\documentclass[bibyear]{aa} 

\usepackage{graphicx}
\usepackage{txfonts}
\begin{document}

\def\mincir{\raise -2.truept\hbox{\rlap{\hbox{$\sim$}}\raise5.truept \hbox{$<$}\ }}
\def\mincireq{\hbox{\raise0.5ex\hbox{$<\lower1.06ex\hbox{$\kern-1.07em{\sim}$}$}}}
\def\magcir{\raise-2.truept\hbox{\rlap{\hbox{$\sim$}}\raise5.truept \hbox{$>$}\ }}

\title{
Non-thermal emission in M31 and M33}

\titlerunning{Relativistic particles in M31 and M33}


   \author{Massimo Persic,
          \inst{1}
          Yoel Rephaeli\inst{2}
          \and
          Riccardo Rando
          \inst{3}
          }

   \institute{
        INAF-Padova Astronomical Observatory, vicolo dell'Osservatorio 5, I-35122 Padova, Italy \\
        INFN-Trieste, via A.\,Valerio 2, I-34127 Trieste, Italy \\
              \email{massimo.persic@inaf.it}
        \and
        School of Physics \& Astronomy, Tel Aviv University, Tel Aviv 69978, Israel \\
        Center for Astrophysics and Space Sciences, University of California at San Diego, La Jolla, CA 92093, USA \\
             \email{yoelr@wise.tau.ac.il}
        \and
        Department of Physics \& Astronomy, University of Padova, via Marzolo 8, 35131 Padova, Italy \\
        Center of Studies and Activities for Space, University of Padova, via Venezia 15, 35131 Padova, Italy \\
        INFN-Padova, via Marzolo 8, 35131 Padova, Italy \\
              \email{riccardo.rando@unipd.it}
             }


 
  \abstract
{Spiral galaxies M31 and M33 are among the $\gamma$-ray sources detected by the {\it Fermi} Large Area Telescope (LAT).}
{We aim to model the broadband non-thermal emission of the central region of M31 (a LAT point source) and of the disk of M33 (a LAT 
extended source),{ as part of our continued survey of non-thermal properties of local galaxies that includes the Magellanic Clouds 
(Persic \& Rephaeli 2022).}}
{ We analysed the observed emission from the central region of M31 ($R < 5.5$ kpc) and the disk-sized emission from M33 ($R \sim 9$ kpc). 
For each galaxy, we self-consistently modelled the broadband spectral energy distribution of the diffuse non-thermal emission based on 
published radio and $\gamma$-ray data. All relevant radiative processes involving relativistic and thermal electrons (synchrotron, Compton 
scattering, bremsstrahlung, and free-free emission and absorption), along with relativistic protons ($\pi^0$ decay following interaction 
with thermal protons), were considered, using exact emissivity formulae. We also used the Fermi/LAT-validated $\gamma$-ray emissivities 
for pulsars. } 
{Joint spectral analyses of the emission from the central region of M31 and the extended disk of M33 indicate that the radio emission is 
composed of both primary and secondary electron synchrotron and thermal bremsstrahlung, whereas the $\gamma$-ray emission may be explained 
as a combination of diffuse pionic, pulsar, and nuclear-BH--related emissions in M31 and plain diffuse pionic emission (with an average 
proton energy density of 0.5 eV cm$^{-3}$) in M33.}
{The observed $\gamma$-ray emission from M33 appears to be mainly hadronic. This situation is similar to other local galaxies, namely, 
the Magellanic Clouds. In contrast, we have found suggestions of a more complex situation in the central region of M31, whose 
emission could be an admixture of pulsar emission and hadronic emission, with the latter possibly originating from both the disk and 
the vicinity of the nuclear black hole. The alternative modelling of the spectra of M31 and M33 is motivated by the different hydrogen 
distribution in the two galaxies: The hydrogen deficiency in the central region of M31 partially unveils emissions from the nuclear BH 
and the pulsar population in the bulge and inner disk. If this were to be the case in M33 as well, these emissions would be outshined 
by diffuse pionic emission originating within the flat central-peak gas distribution in M33.}


   \maketitle
%

\section{Introduction}

Non-thermal phenomena are mainly probed by the radiative yields of relativistic electrons and protons (hereafter simply referred to as 
electrons and protons, unless noted otherwise). Measurements of the emission spectral energy distribution (SED) provide direct insight 
on the particle energy spectra and on magnetic and radiation fields in star-forming environments.

In a previous paper (Persic \& Rephaeli 2022: Paper I), we examined non-thermal emission in the Magellanic Clouds, satellite galaxies 
of the Milky Way. Adopting a one-zone model for the extended disk emission, we self-consistently modelled disk non-thermal emission at 
radio and $\gamma$-ray frequencies, interpreting radio data as synchrotron plus thermal bremsstrahlung, and $\gamma$-ray data as neutral-pion 
decay (following energetic proton interactions in the ambient gas) plus leptonic emission.

In this follow-up paper on the non-thermal properties of star-forming galaxies, we consider the two Local Group member galaxies M31 and 
M33 -- at distances of 0.78 and 0.85 Mpc, respectively.
\footnote{
For our purposes, the source regions of the two galaxies are modelled as cylinders of radius and height 
$R_s = 5.45$ kpc and $h_s = 0.24$ kpc (Brinks \& Burton 1984) for M31, and $R_s = 8.7$ kpc and $h_s = 
0.2$ kpc (Kam et al. 2015; Monnet 1971) for M33. 
}
As it was for the Magellanic Clouds in Paper I, our main objective in this study is to determine relativistic electron and proton spectra 
in the disks of M31 and M33 by spectral modelling of their non-thermal emission in all the relevant energy ranges accessible to observations. 
In this paper, too, we focus on the extended disk emission of each galaxy because this emission traces the mean galactic properties of non-thermal 
particles and magnetic fields. Their integrated radio spectra have been measured long ago (Beck 2000 and references therein; Dennison et al. 
1975), whereas it is only recently that $\gamma$-ray data have become available (Abdo et al. 2010, Xi et al. 2020) -- but there are no 
published measurements 
of extended non-thermal X-ray emission. Adopting a one-zone model for the extended disk emission we calculate the radiative yields of 
electrons and protons and contrast these with available radio and $\gamma$-ray data. The latter are fully specified in Table\,1 (M31) and 
Table\,2 (M33). 
 
In Section 2, we review published observations of broadband non-thermal interstellar emission from M31 and M33. In Section 3, we review the 
radiation fields permeating the two galaxies. In Section 4, we describe the SED modelling procedure. In Sections 5 and 6, we present and 
compare our SED models to the data for M31 and M33, respectively. We summarise our results in Section 7.

\section{Observations of extended emission}

M31 and M33 have been extensively observed over a wide range of radio--microwave, (soft) X-ray, and $\gamma$-ray bands; point-source, 
extended small- and large-scale emission has been detected over a wide spectral range. Their detected disk X-ray diffuse emission is 
thermal from hot ($T_e \sim 10^6\,\mathrm{K}$) ionised plasma. It is relevant to our SED analysis as it contributes to the ambient 
proton density that enters the calculation of the pionic $\gamma$-ray spectrum. Both galaxies were searched for $\gamma$-ray emission 
early on in the {\it Fermi}-LAT mission (Abdo et al. 2010): M31 first, and later M33, were detected as marginally extended sources 
(Abdo et al. 2010; Xi et al. 2020). 

Emission from both galaxies is consistent with an empirical, quasi-linear $L_{>0.1\, {\rm GeV}}$ versus $L_{8-1000\,\mu{\rm m}}$ 
correlation established for nearby star-forming galaxies (Ackermann et al. 2012), suggesting a SF-related origin of their $\gamma$-ray 
emission. In this section, we review the observations most relevant to non-thermal emission, referring for details to the cited papers.

\subsection{M31}

\noindent
$\bullet$ {\it Radio.} Battistelli et al. (2019) deduced an overall sky-integrated
\footnote{ 
RA $\times$ DEC $= 2^{\circ}.4 \times 3^{\circ}.1 = 7.4$ deg$^2$.
}
density spectrum of M31 from the radio to the infrared (IR), based on a compilation of  Sardinia Radio Telescope (SRT) 6.7 GHz 
measurements, newly analysed {\it Wilkinson} Microwave Anisotropy Probe ({\it W}MAP) and {\it Planck} data, as well as other 
ancillary data -- after subtracting point sources at $\leq$3$\sigma$ above noise estimation. Decomposing the radio-IR SED into 
synchrotron, thermal free-free, thermal dust, and the 'anomalous' microwave emission (AME) arising from electric dipole radiation 
from spinning dust grains in ordinary interstellar conditions (Draine \& Lazarian 1998a,b; see Dickinson et al. 2018 for a 
review), Battistelli et al. (2019) found that at $\nu < 10$ GHz the SED is dominated by synchrotron emission with a spectral 
index $\alpha = 1.10^{\scriptscriptstyle +0.10}_{\scriptscriptstyle -0.08}$ (in agreement with Berkhuijsen et al. 2003) combined 
with a subdominant $\propto \nu^{-0.1}$ free-free emission at the highest frequencies. At $\nu \sim 10$ GHz AME becomes comparable 
to synchrotron and free-free emissions, at 20 GHz $< \nu <$ 50 GHz it dominates the emission, and at $\nu > 100$ GHz, thermal dust 
emission is virtually the only emission component. Thus our analysis of non-thermal emission will use the Battistelli et al. (2019) 
data compilation at low frequencies only ($\nu < 10$ GHz; Table 1), where AME and thermal dust emission are negligible.

A peculiar feature characterises the (atomic and molecular) hydrogen distribution in M31. The central galactic region is poor in 
hydrogen, to the extent that gas is nearly absent in the nuclear region. 
\medskip

\noindent
$\bullet$ {\it X-ray.} 
Presence of diffuse emission in the bulge was suggested by Primini et al. (1993) and Supper et al. (1997), based on {\it R\"ontgen 
Satellit} (ROSAT) soft X-ray data after point-source subtraction
\footnote{ 
86 sources in the central $\sim$34$^\prime$ down to $L_{0.2-4 {\rm keV}} = 1.4\,10^{36}$ erg s$^{-1}$ from a 48\,ks 
observation with the High Resolution Imager (HRI; Primini et al. 1993); and 396 sources within 6.3 deg$^2$ in the 
luminosity range to $3\,10^{35} < L_{0.1-2.4 {\rm keV}}/{\rm erg\,s}^{-1} < 2\,10^{38}$ from a 205\,ks observation 
with the Position Sensitive Proportional Counters (PSPC; Supper et al. 1997), respectively. 
}, 
with a luminosity $L_{0.1-2.4 {\rm keV}} < 3\,10^{38}$ erg s$^{-1}$. 
Unresolved emission from the disk was clearly seen in residual maps after subtraction of point sources, instrumental, and extragalactic 
backgrounds (West et al. 1997, based on {\it ROSAT} PSPC and HRI data). Two components were distinguished, a spherical core plus an 
exponential disk with a radial profile similar to that of the optical light (i.e. the surface brightness is described as $I(R) \propto 
e^{-(R/R_d)}$ where $R_d = 5$ kpc is the radial scalelength). The disk diffuse emission makes up $\sim$45\% of the total diffuse emission 
and corresponds to a hot plasma mass of $M_X < 7\, 10^6 M_\odot$. Takahashi et al. (2001: Advanced Satellite for Cosmology and Astrophysics 
[ASCA] data) interpreted spatially averaged (out to galactocentric radii $12^\prime$) spectra as integrated emission ($L_{0.5-10 {\rm keV}} 
= 2.6\,10^{39}$ erg s$^{-1}$) from a population of low-mass X-ray binaries 
\footnote{
Hard (3$-$100 keV) X-ray data the emission shows two components, one broadly peaking at $\sim$5 keV, and another 
extending to $>$100 keV, that are interpreted as arising from NS binaries with, respectively, high/low accretion 
rates and luminosities above/below $\sim$10$^{37}$ erg s$^{-1}$ (Revnitsev et al. 2014: {\it Rossi} X-ray Timing 
Explorer ({\it R}XTE)/Proportional Counter Array (PCA), INTErnational Gamma-Ray Astrophysics Laboratory (INTEGRAL)/Integral Soft 
Gamma-Ray Imager (ISGRI), and {\it Neil Gehrels} Swift Observatory/Burst Alert Telescope (BAT) data).}
plus two diffuse thermal plasma models characterised by $k_BT=0.3$ and 0.9 keV both exhibiting $L_{0.5-10 {\rm keV}} = 2\,10^{38}$ erg 
s$^{-1}$. 

Bogdan \& Gilfanov (2008; {\it Chandra} and XMM-{\it Newton} data) detected soft emission from ionised gas with a temperature of $k_BT 
\sim 0.3$ keV and $M_X \sim 2\, 10^6 M_\odot$, significantly distributed along the minor axis of the galaxy, possibly indicating outward 
flowing gas perpendicular to the disk. The vertical extent of the gas is $\magcir$2.5 kpc, as suggested by the shadow (suppressed emission) 
seen on the 10 kpc radius star-forming ring at $\sim10^\prime$ from the centre. 

This review of the X-ray literature on M31 has revealed no non-thermal X-ray flux that, interpreted as Comptonised cosmic 
microwave background (CMB) radiation, could have been used to calibrate the relativistic electron spectrum (e.g. Persic 
\& Rephaeli 2019a). The detected X-ray emission is thermal from hot ($T_e \sim 10^6\,\mathrm{K}$) ionised plasma, whose density 
contributes to the ambient density in the p--p interactions that lead to the production of neutral pions ($\pi^0$, which quickly decay 
into $\gamma$\,rays) and charged pions ($\pi^\pm$, which ultimately produce secondary electrons and positrons) as well as non-thermal 
bremsstrahlung.
\medskip

\noindent
$\bullet$ {\it $\gamma$ ray.}
M31 was detected at 5$\sigma$ significance with $\mincir$2\,yr of {\it Fermi}-LAT data as a marginally extended source in the 0.2$-$20 
GeV band (Abdo et al. 2010), and then again (at $\mincir$10$\sigma$) with $>$7\,yr of LAT Pass 8 data as an extended source in the energy 
range 0.1$-$100 GeV (Ackermann et al. 2017)
\footnote{
Based on 33 months of High Altitude Water Cherenkov (HAWC) telescope data, an upper limit was set on the 1$-$100 TeV flux from the galactic 
disk of M31 (Albert et al. 2020).}. 
The $\gamma$-ray brightness distribution was found to be consistent with a $0^\circ.4$-radius uniform-brightness disk and no offset from 
the galaxy centre, with a spectrum compatible with either a power-law (PL) profile (index $2.4 \pm 0.1$) or a hump -- more likely to 
originate from $\pi^0$ decay than from Compton scattering of starlight. The emission appeared to originate from the inner regions of the 
galaxy, and seemed not to be correlated with regions rich in gas or ongoing star-formation activity, suggesting (according to Ackermann 
et al. 2017) a non-interstellar origin unless the energetic protons originated in previous episodes of star formation. On the other hand 
McDaniel et al. (2019), using multi-frequency data, suggested that the emission most likely has a lepto-hadronic (LH) origin, namely, a 
combination of pionic and Compton scattering of primary and secondary electrons off ambient radiation fields, with deduced parameter 
values consistent with previous studies of M31 and cosmic-ray physics. We note also that Di Mauro et al. (2019) argued against dark-matter 
decay as a source for the measured emission.

In a spectro-morphological analysis employing templates for the distribution of the stellar mass of the galaxy and updated astrophysical 
foreground models for its sky region, Zimmer et al. (2022) constructed maps of the old stellar population in the bulge that mimicked the 
distribution of a population of millisecond pulsars (msPSR). Analysing LAT data using such templates they obtained a 5.4$\sigma$ detection, 
which is comparable with the detection by Ackermann et al. (2017)  using a disk template. Zimmer et al. (2022) argued that $\sim$10$^6$ 
unresolved msPSR could account for the measured $\gamma$-ray luminosity and spectrum; the latter then interpreted as arising from Compton 
upscattering of ambient photons by electrons in the bulge that were accelerated by msPSRs in the disk. Eckner et al. (2018) compared the 
extended emission detected by Ackermann et al. (2017) with that of a msPSR population in M31 obtained from the one detected in the local 
Milky Way disk via rescaling by the respective stellar masses of the systems: with no free parameters, they estimated a contribution of 
$\sim$1/4 of the M31 emission, as well as the energetics and morphology of the Galactic Center 'GeV excess'. Fragione et al. (2019) 
estimated that msPSR in old globular clusters (that were likely dragged to the centre by dynamical friction) may contribute $\sim$1/8 of 
the emission deduced by Ackermann et al. (2017).

Using 7.6\,yr of LAT data, Karwin et al. (2019) carried out a detailed study of the 1$-$100 GeV emission toward M31's outer halo with a 
total field radius of $60^\circ$ centred on M31. Accounting for foreground $\gamma$-ray emission from the Milky Way, they suggested the 
existence of an extended excess unrelated to the Milky Way and having a total radial extent of $\mincir$200 kpc from the centre of M31. 
While not ruling out an extended cosmic-ray halo underlying such purported emission, citing GALPROP predictions Karwin et al. (2019) 
argued against it based on the radial extent, spectral shape, and intensity of the large-scale emission. However, Do et al. (2021) and 
Recchia et al. (2021) pointed out that the emission radial extent and intensity strongly depend on assumptions on cosmic-ray diffusion 
outside the plane and in the halo of M31 and that the spectral shape is affected by foreground Milky Way emission and by the intrinsic 
weakness and limited statistics of the signal, so the emission out to $\mincir$200 kpc may indeed originate from an extended cosmic-ray 
leptonic and/or hadronic halo if cosmic-ray propagation is more effective than assumed by Karwin et al. (2019). Based on a model-dependent 
approach, Zhang et al. (2021) used the purported halo emission to estimate the total baryon mass contained in the circumgalactic medium 
within M31's the $\sim$250 kpc virial radius of M31 and suggested it could account for $<$30\% of the missing (with respect to the 
cosmic abundance) baryons.

Most recently Xing et al. (2023) carried out an analysis of the expanded $\sim$14\,yr LAT database, finding that the extended source 
claimed by Ackermann et al. (2017) actually consists of two separate regions: a central emission region, with a log-parabola spectrum, 
coincident with the optical centre of the galaxy, and a second region, with a PL spectrum, $0^\circ.4$ southeast of the centre (note: 
it is unclear whether the latter is a background extragalactic source or a local source associated with M31). Xing et al. (2023) argued 
that the newly revealed point-like nature of the central source necessitates a revised interpretation since the Ackermann et al. (2017) 
conclusion was based on the assumption that the central emission has an extended distribution (cosmic rays, pulsars). 

This review of the $\gamma$-ray literature on M31 has shown that previous $\gamma$-ray observations were affected by source confusion. This 
motivates our new interpretation of the M31 central emission in Section 5.

\subsection{M33}

\noindent
$\bullet$ {\it Radio.}
A study of the large-scale radio emission over the wide frequency range 21--10,700 MHz was carried out by Israel et al. (1992), 
conflating original data (21--610 MHz) and published data (178--10,700 MHz). In light of the $18^\prime-1^\prime$ beamsize range 
and point-source subtraction procedure, this database (Table 2) is still the most suitable for our analysis of the galaxy-wide 
integrated radio properties of M33
\footnote{ 
The optical size of M33, as measured by the 25-mag-arsec$^{-2}$ radius, is $R_{25}=35.4^\prime$; de Vaucouleurs et al. 1991).
}
For example, the main focus of the recent deep 1.5 and 5 GHz { {\it Jansky} Very Large Array} ({\it J}VLA) survey of this galaxy 
(White et al. 2019) is a detailed mapping of point sources.
 
Israel et al. (1992) used the method from Baars et al. (1977)  to convert instrumental intensities to flux densities. The $<$100 
MHz flux densities were obtained with the University of Maryland's (defunct) Clark Lake Radio Observatory: their uncertainties 
mostly stem from the low surface brightness of the radio disk and the relatively high confusion level for a large beam at low 
frequencies. The 151.5 MHz flux density was obtained from several observations carried out with Cambridge University's Mullard 
Radio Astronomy Observatory, and the 327 and 610 MHz flux densities were obtained from maps made with the Westerbork Synthesis 
Radio Telescope. 

Virtually all the M33 emission analysed by Israel et al. (1992) is contained in a box of the following size: $45^\prime \times 
85^\prime$. These authors noted a spectral turnover at $\sim$750 MHz, and proposed a double-PL (2PL) spectral model with low/high 
index of 0.18/0.86 (compatible with previous works in the literature, based on much sparser data). To explain this spectral shape 
they suggested free-free absorption of the synchrotron emission by a cool ($\leq 1000\,\mathrm{K}$) ionised plasma to be a very 
likely possibility.

The circularly integrated HI surface distribution shows a broad ($R \mincir 6$\,kpc) flat central peak. At larger radii the 
distribution declines steeply and cuts off at $\sim$10\,kpc (Wright et al. 1972).
\medskip

\noindent
$\bullet$ {\it X-ray.} 
Diffuse emission in M33 was first suggested based on {\it Einstein} Imaging Proportional Counter (IPC) and HRI data. In addition 
to a dominant nuclear source and 17 point-like sources located mostly in the spiral arms (Long et al. 1981; Markert \& Rallis 1983), 
Trinchieri et al. (1988) identified a diffuse low-surface--brightness emission from the galactic plane to which source confusion 
and hot ($kT < 1$ keV) gas could contribute. Using deeper ROSAT HRI data, Schulman \& Bregman (1995) detected more point sources 
above their data sensitivity threshold and noted that the (residual) extended X-ray disk emission (with a luminosity of $L_{0.1-2.4 
{\rm keV}} \sim 10^{39}$ erg s$^{-1}$) is not due to (additional) point sources.

Finally, Long et al. (1996) used deep (50.4 ks) {\it ROSAT} PSPC data to expand the list of individual sources to 57 
\footnote{
Combining all {\it ROSAT} observations Haberl \& Pietsch (2001) conclusively 
found 184 sources within $50^{\prime\prime}$ of the nucleus.
}
and confirmed the diffuse disk emission (possibly tracing the spiral arms) to be of thermal ($kT \sim 0.4$ keV) origin. Using 
{\it XMM-Newton} data $\magcir$10 times deeper than earlier {\it ROSAT} observations, Pietsch et al. (2004) and Misanovic et 
al. (2006) detected 350 sources in the galaxy sky area (of which $\sim$200 are in M33) and again found evidence of a diffuse 
emission. A subsequent very deep (1.4 Ms) {\it Chandra} Advanced CCD Imaging Spectrometer (ACIS) survey of M33 (Plucinsky et 
al. 2008) revealed soft diffuse emission from the large-scale hot ionised plasma and specific extended regions (e.g. the 
star-forming HII regions NGC\,604, with $kT \sim 0.2$ keV; or IC\,131 of enigmatic spectral interpretation: T\"ullmann et al. 
2009). 

As in the case of M31, the conclusion of this survey of the X-ray literature on M33 is that only thermal X-ray emission from hot 
plasma has been detected thus far. The plasma density is (obviously) used in the calculation of the outcome of p--p interaction 
and non-thermal bremsstrahlung.
\medskip

\noindent
$\bullet$ {\it $\gamma$ ray.}
M33 was not detected in the $\mincir$2\,yr of {\it Fermi}-LAT data (Abdo et al. 2010), nor in the $>$7\,yr of data (Ackermann et 
al. 2017). Karwin et al. (2019) saw positive residual emission in the direction to M33 while examining the $\gamma$-ray emission 
from the outer halo of M31. Di Mauro et al. (2019) claimed a point-like source detection of M33; however, this was challenged by 
Xi et al. (2020) on grounds that the Di Mauro et al. background model had missed three background sources positionally close to 
M33; if unaccounted for, these would be attributed to M33. Finally, using a state-of-the-art background emission model and 11.4\,yr 
of LAT data, Xi et al. (2020) claimed a detection of M33: the LAT image shows an extended signal peaking on the northeast region 
of the galaxy, where its most massive HII region, NGC\,604, is located. We use the Xi et al. (2020) data (Table 2).

\begin{table*}
\caption[] {M31 radio and $\gamma$-ray data.}
\centering 
\begin{tabular}{ l  l  l  l  l  l  l}
\hline
\hline
\noalign{\smallskip}
Frequency    &        Energy flux                                                 & Reference                                  & & Frequency          &   Energy flux                                                     & Reference                       \\
log($\nu$/Hz)&   log(erg\,cm$^{-2}$s$^{-1}$) &                                                                                 & & log($\nu$/Hz)      & log(erg\,cm$^{-2}$s$^{-1}$)                                       &                                 \\
\noalign{\smallskip}
\hline
\noalign{\smallskip}
7.538        &  $-13.498^{\scriptscriptstyle +0.074}_{\scriptscriptstyle -0.089}$ &{\small Dwarakanath \& Udaya Shankar (1990)}& &  9.841             & $-13.083^{\scriptscriptstyle +0.031}_{\scriptscriptstyle -0.034}$ &{\small Battistelli et al. (2019)}\\
8.611        &  $-13.124^{\scriptscriptstyle +0.036}_{\scriptscriptstyle -0.040}$ &{\small Haslam et al. (1981)}               & & $23.154 \pm 0.151$ & $-12.180^{\scriptscriptstyle +0.149}_{\scriptscriptstyle -0.228}$  &{\small Xing et al. (2023)}   \\
9.105        &  $-13.160^{\scriptscriptstyle +0.032}_{\scriptscriptstyle -0.034}$ &{\small Melis et al. (2018)}                & & $23.462 \pm 0.151$ & $-12.387^{\scriptscriptstyle +0.128}_{\scriptscriptstyle -0.181}$  &{\small Xing et al. (2023)}     \\
9.152        &  $-13.125^{\scriptscriptstyle +0.032}_{\scriptscriptstyle -0.035}$ &{\small Reich et al. (2001)}                & & $23.770 \pm 0.151$ & $-12.420^{\scriptscriptstyle +0.110}_{\scriptscriptstyle -0.148}$  &{\small Xing et al. (2023)}     \\
9.158        &  $-13.121^{\scriptscriptstyle +0.030}_{\scriptscriptstyle -0.033}$ &{\small Melis et al. (2018)}                & & $24.078 \pm 0.151$ & $-12.699^{\scriptscriptstyle +0.161}_{\scriptscriptstyle -0.260}$  &{\small Xing et al. (2023)}   \\
9.800        &  $-13.118^{\scriptscriptstyle +0.032}_{\scriptscriptstyle -0.031}$ &{\small Battistelli et al. (2019)}          & & $24.386 \pm 0.151$ & $-12.721^{\scriptscriptstyle +0.213}_{\scriptscriptstyle -0.434}$  &{\small Xing et al. (2023)}     \\                                                                                                  
\noalign{\smallskip}
\hline
\hline
\end{tabular}
\end{table*}

\begin{table*}
\caption[] {M33 radio and $\gamma$-ray data.}
\centering 
\begin{tabular}{ l  l  l  l  l  l  l}
\hline
\hline
\noalign{\smallskip}
Frequency    &        Energy flux  & Reference                          & & Frequency            &   Energy flux           & Reference                          \\
log($\nu$/Hz)& $10^{-14}$erg\,cm$^{-2}$s$^{-1}$&                        & &log($\nu$/Hz)         & $10^{-14}$erg\,cm$^{-2}$s$^{-1}$ &                           \\
\noalign{\smallskip}
\hline
\noalign{\smallskip}
7.330        &  $0.150 \pm 0.064$  &{\small Israel et al. (1992)}       & &   9.149              &  $4.508 \pm 0.705$      &{\small Dennison et al. (1975)}     \\
7.408        &  $0.307 \pm 0.103$  &{\small Israel et al. (1992)}       & &   9.151              &  $4.508 \pm 0.084$      &{\small Hutchmeier (1975)}          \\
7.490        &  $0.185 - 0.371$    &{\small Israel et al. (1992)}       & &   9.152              &  $4.260 \pm 0.568$      &{\small Buczilowski (1988)}         \\
7.760        &  $0.431 \pm 0.144$  &{\small Israel et al. (1992)}       & &   9.236              &  $4.644 \pm 0.459$      &{\small Buczilowski (1988)}         \\
7.848        &  $0.211 - 0.845$    &{\small Israel et al. (1992)}       & &   9.431              &  $6.468 \pm 2.695$      &{\small Dennison et al. (1975)}     \\
8.180        &  $0.848 \pm 0.273$  &{\small Israel et al. (1992)}       & &   9.432              &  $4.593 \pm 0.459$      &{\small Buczilowski (1988)}         \\
8.514        &  $1.844 \pm 0.261$  &{\small Israel et al. (1992)}       & &   9.677              &  $5.225 \pm 0.808$      &{\small Buczilowski (1988)}         \\
8.785        &  $2.682 \pm 0.366$  &{\small Israel et al. (1992)}       & &   9.686              &  $6.790 \pm 1.455$      &{\small von Kap-herr et al. (1978)} \\
8.250        &  $1.780 \pm 0.712$  &{\small Leslie (1960)}              & &  10.029              &  $5.885 \pm 1.605^\star$&{\small Buczilowski (1988)}         \\
8.502        &  $2.449 \pm 0.382$  &{\small Terzian \& Pankonin (1972)} & &                      &  --------------------   &                                    \\
8.611        &  $2.448 \pm 0.245$  &{\small Braccesi et al. (1967)}     & &                      & $10^{-13}$erg\,cm$^{-2}$s$^{-1}$ &                           \\
8.782        &  $3.636 \pm 0.364$  &{\small Terzian \& Pankonin (1972)} & &                      & --------------------    &                                    \\
8.875        &  $4.950 \pm 1.350$  &{\small Venugopal (1963)}           & & $22.790 \pm 0.407$   &  $<7.5$                 &{\small Xi et al. (2020)}           \\
8.875        &  $3.600 \pm 1.875$  &{\small De Jong (1965)}             & & $23.286 \pm 0.704$   &  $2.034 \pm 0.066$      &{\small Xi et al. (2020)}           \\
8.925        &  $4.547 \pm 1.010$  &{\small Buczilowski (1988)}         & & $24.082 \pm 0.704$   &  $1.556 \pm 0.056$      &{\small Xi et al. (2020)}           \\
9.146        &  $5.740 \pm 2.100$  &{\small Venugopal (1963)}           & & $24.860 \pm 0.704$   &  $1.045 \pm 0.068$      &{\small Xi et al. (2020)}           \\
9.149        &  $4.512 \pm 2.397$  &{\small de Jong (1965)}             & & $25.684 \pm 0.704$   &  $<10$                  &{\small Xi et al. (2020)}           \\

\noalign{\smallskip}
\hline
\hline
\end{tabular}
\flushleft
$^\star${\small Considered to be a lower limit (Israel et al. 1992).}
\smallskip
\end{table*}

\section{Radiation fields}

A reasonably precise determination of the ambient radiation field is needed to predict the level of $\gamma$-ray emission from Compton 
scattering of {target photons by} the radio-emitting electrons (and positrons). The total radiation field includes cosmic (background) 
and local (foreground) components. 

Relevant cosmic radiation fields include the CMB, a pure Planckian with a local temperature of $T_{\rm CMB} = 2.735\,\mathrm{K}$ and 
energy density of $u_{\rm CMB} = 0.25$ eV cm$^{-3}$, and the extragalactic background light (EBL). The latter originates from direct 
and dust-reprocessed starlight integrated over the star formation history the Universe, with two respective peaks that are referred 
to as the cosmic optical background (at $\sim$1\,$\mu$m), and cosmic infrared background (at $\sim$100\,$\mu$m). The two peaks are 
described as diluted Planckians, characterised by a temperature, $T,$ and a dilution factor, $C_{\rm dil}$. The latter is the ratio 
of the actual energy density, $u$, to the energy density of an undiluted blackbody at the same temperature, $T$ (i.e. $u = C_{\rm 
dil} \frac{4}{c} \sigma_{\rm SB} T^4$, where $\sigma_{\rm SB}$ is the Stefan-Boltzmann constant). The widely used EBL model, based 
on galaxy counts in several spectral bands (Franceschini \& Rodighiero 2017), can be numerically approximated as a combination of 
diluted Planckians, 
\begin{eqnarray}
\lefteqn{
n_{\rm EBL}(\epsilon) ~=~ \sum_{j=1}^8 A_j \,\frac{8 \pi}{h^3c^3} \, \frac{\epsilon^2}{e^{\epsilon/k_B T_j}-1} \hspace{0.5cm}  
{\rm cm^{-3}~ erg^{-1}}, }
\label{eq:EBL}
\end{eqnarray}
where $A_j$ and $T_j$ are suitable dilution factors and temperatures. 
\footnote{ 
$A_1=10^{-5.629}$, $T_1=29\,\mathrm{K}$; 
$A_2=10^{-8.522}$, $T_2=96.7\,\mathrm{K}$; 
$A_3=10^{-10.249}$, $T_3=223\,\mathrm{K}$; 
$A_4=10^{-12.027}$, $T_4=580\,\mathrm{K}$;  
$A_5=10^{-13.726}$, $T_5=2900\,\mathrm{K}$; 
$A_6=10^{-15.027}$, $T_6=4350\,\mathrm{K}$; 
$A_7=10^{-16.404}$, $T_7=5800\,\mathrm{K}$;
$A_8=10^{-17.027}$, $T_8=11600\,\mathrm{K}$.
}

Local radiation fields in the two galaxies (galactic foregound light, GFL) arise from their stellar populations. Similar to 
the EBL by shape and origin, the GFL is dominated by two thermal humps, IR and optical. Full-band luminosities of these thermal 
components are needed to determine $n(\epsilon)$, the spectral distributions of the photons to be Compton scattered (shown in 
Eq.\,\ref{eq:IC_emissivity2}). We compute the total IR ($8-1000\,\mu$m) luminosities using InfraRed Astronomical Satellite ({\it 
IRAS}) flux densities at 12$\mu$m, 25$\mu$m, 60$\mu$m, and 100$\mu$m (sampling the IR hump), from the relation $f_{\rm IR}=1.8 
\cdot 10^{-11} (13.48\,f_{12\,\mu{\rm m}} + 5.16\,f_{25\,\mu{\rm m}} + 2.58\,f_{60\,\mu{\rm m}} + f_{100\,\mu{\rm m}})$\, erg\,
cm$^{-2}$s$^{-1}$ (Sanders \& Mirabel 1996). The total optical luminosities are computed from narrow-band luminosities (H band 
for M31, B band for M33) by applying suitable bolometric corrections that restore the full-band luminosities. The corresponding 
energy densities, $u_k=L_k/\left[2\pi c R_s(R_s+h_s)\right]$, with $k$ denoting either IR or optical, are: 
\smallskip

\noindent
{\it (a)} M31: The bolometric IR luminosity, $L_{\rm IR} = 1.2\, 10^{43}$\,erg\,s$^{-1}$, is computed from the {\it IRAS} flux 
densities reported in Table 3. The bolometric optical luminosity, $L_{\rm opt} = 1.2\, 10^{45}$\,erg\,s$^{-1}$, is computed from 
the narrow-band 3.6$\mu$m IRAC luminosity, $L_{\rm IRAC} = 1.17\, 10^{11} L_\odot$ (Courteau et al. 2011), multiplied by the 
bolometric correction, $f_{\rm bol} = 2.7$ (Rosenfield et al. 2012). Our region of interest is within $R_{\star} = 1.09 \, R_d$, 
within which only 30\% of the galaxy-wide emission is produced and is relevant to our GFL computation. 
\footnote{
The bulge contribution is very minor (Soifer et al. 1986). 
}
Given the relative thinness of the radio disk, only a negligible fraction of the radiation emitted at $R > R_{\star}$ enters the 
inner region. We obtain $u_{\rm IR}= 0.04$ eV\,cm$^{-3}$ and $u_{\rm opt}= 4$ eV\,cm$^{-3}$. The IR and optical hump temperatures 
reported in Table 3 imply dilution factors for the M31-sourced photon fields of $c_{\rm IR} = 10^{-4.925}$, and $C_{\rm opt} = 
10^{-10.917}$; 
\smallskip

\noindent
{\it (b)} M33: The bolometric IR luminosity, $L_{\rm IR} = 4.8\, 10^{42}$\,erg\,s$^{-1}$, is computed from the data reported 
in Table 3. The total (bolometric) optical luminosity, $L_{\rm opt} = 3.1\, 10^{43}$\,erg\,s$^{-1}$, is computed from the 
de-reddened blue magnitude and color index, $B_T^0=5.92$ and $(B-V)_0 \sim 0.47$ (Rice 1988), by applying Buzzoni et al.'s 
(2006) bolometric correction as described in Persic \& Rephaeli (2019a). Thus, $u_{\rm IR}= 0.022$ eV\,cm$^{-3}$ and $u_{\rm opt} 
= 0.14$ eV\,cm$^{-3}$. The IR and optical hump temperatures (Table 3) imply $c_{\rm IR} = 10^{-5.267}$, and $C_{\rm opt} = 
10^{-12.436}$.

\begin{table}

\caption[] {M31 and M33: stellar-population emission parameters.}
\centering 
\begin{tabular}{ l  l  l  l  l  l  l}
\hline
\hline
\noalign{\smallskip}
Galaxy  & $f_{12\mu{\rm m}}$ &   $f_{25\mu{\rm m}}$  & $f_{60\mu{\rm m}}$ & $f_{100\mu{\rm m}}$ & $T_{\rm IR}$ & $T_{\rm opt}$ \\
        &       Jy           &         Jy            &           Jy       &          Jy         & $\mathrm{K}$ & $\mathrm{K}$  \\
\noalign{\smallskip}
\hline
\noalign{\smallskip}
M31$^a$ &         170        &           220         &          690       &         3800        &        29    &   2900        \\
M33$^b$ &          33        &            40         &          420       &         1260        &        30    &   3000        \\

\noalign{\smallskip}
\hline\end{tabular}
\smallskip

\noindent
{\small $^a$ {\it IRAS} flux densities from Habing et al. (1984). Temperatures $T_{\rm IR}$ and $T_{\rm opt}$ estimated using Wien law 
from the IR peak at 160\,$\mu$m (Montalto et al. 2009; Viaene et al. 2014) and the optical peak at 1\,$\mu$m (Viaene et al. 2014). }
\smallskip

\noindent
{\small $^b$ {\it IRAS} flux densities from Rice et al. (1990). Temperatures $T_{\rm IR}$ and $T_{\rm opt}$ estimated using Wien law 
from the IR peak at 100\,$\mu$m and the optical peak at 1.1\,$\mu$m (Rice et al. 1990). }
\smallskip
\end{table}

\section{SED models}

As mentioned, the main objective of this study is to determine the spectra of interstellar relativistic electrons and protons in the 
disks of M31 and M33 by a spectral modelling { of the} non-thermal emission in all the relevant energy ranges accessible to observations. 
In general, the particle, gas density, magnetic, and radiation field distributions vary significantly across galaxy disk, obviating the 
need for a spectro-spatial treatment: indeed, a detailed modelling approach based on a solution to the diffusion-advection equation has 
been applied in the study of the Galaxy and nearby galaxies where the spatial profiles of key quantities, such as the particle acceleration 
sources, gas density, magnetic field, and generally also the particle diffusion coefficient may be specified (e.g. Strong et al. 2007, 2010; 
Jones 1978; Taillet \& Maurin 2003; Heesen 2021). However, if the aim is gaining insight on spatially-averaged non-thermal properties in the 
disk, then a parameter-intensive spectro-spatial modelling approach is not warranted. While 'top-down' diffusion-based modelling of the 
non-thermal emission in the disks and halos of galaxies is possible (Syrovatskii et al. 1959; Wallace et al. 1980; Rephaeli \& Sadeh 2019), 
here we adopted a 'bottom-up' approach. Thus, we started from the non-thermal yields of relativistic particles and we deduced their 
steady-state spectra, making it is possible (in principle) to deduce the injection spectrum.

Radio emission in disks of spiral galaxies is mostly (electron) synchrotron in a disordered magnetic field whose mean total value, $B,$ is 
taken to be spatially uniform in the regions of interest 
\footnote{ 
This assumption is realistic for the two galaxies considered in this paper (M31: Table 1 of Fletcher 
et al. 2004; M33: Fig.\,9 of Tabatabaei et al. 2008) and it also appears to be valid overall (Beck 2016).
}
plus thermal free-free from a warm ($\sim$10$^4\,\mathrm{K}$) ionised plasma (Spitzer 1978). This emission may be absorbed, at low 
frequencies, by cold ($\mincir 10^3\,\mathrm{K}$) plasma (Israel \& Mahoney 1980; Oster 1961). 
\footnote{
For a review on the multiphase insterstellar medium, see Cox 2005.
}
A significant proton component contributes additional radio emission by secondary $e^{\pm}$ produced by $\pi^{\pm}$ decays and $\gamma$-ray 
emission from $\pi^0$ decay (Stecker 1971). In addition, $\gamma$-ray emission is produced by electron scattering off photons of local 
and background radiation fields (Blumenthal \& Gould 1970). The calculations of the emissivities from all these processes are standard. 

We just briefly expand here on thermal absorption of radio emission since it was not discussed in Paper I. A key feature in modelling the 
radio data is accounting for free-free absorption of non-thermal synchrotron by a warm thermal ionised plasma, as suggested by a high value 
of the H$\alpha$ emission measure, $EM \equiv \int n_e^2 {\rm d}l$. This was originally pointed out for a sample of highly inclined spiral 
galaxies, including M33 and M31, based on the low ratio between the observed radio intensities and the intensities extrapolated from 
measurements at higher frequencies, with the ratio being smallest for edge-on galaxies (Israel \& Mahoney 1980). The frequency-dependent 
optical depth for free-free absorption by thermal plasma is (Israel \& Mahoney 1980): 
\begin{eqnarray}
\lefteqn{
\tau_{ff}(\nu) ~=~ 3.014\, 10^{-2} \, T_e^{-1.5} \, \nu^{-2} \, {\rm EM} \,\, \phi(\nu, T_e),  }
\label{eq:tau1}
\end{eqnarray}
whereby 
\begin{eqnarray}
\lefteqn{
\phi(\nu, T_e) ~=~ {\rm ln}(4.955\, 10^{-2} \, \nu^{-1}) \, + \, 1.5\, {\rm ln}T_e,  }
\label{eq:tau2}
\end{eqnarray}
where the electron temperature $T_e$ is in $\mathrm{K}$, the frequency $\nu$ is in GHz, and the emission measure of the absorbing gas, 
EM, is in cm$^{-6}$ pc. We assume  that the radio emission and absorption are well mixed throughout the galaxy, so the emerging 
intensity is $I_\nu \propto j_\nu (1-e^{-\tau_{ff}(\nu)})/\tau_{ff}(\nu)$, where $j_\nu$ is the (synchrotron plus thermal free-free) 
spectral emissivity and $\tau_{ff}(\nu)$ curves the spectrum at low frequencies.

We assume the particle spectral distributions to be time-independent and locally isotropic; then we have:\ 
\smallskip

\noindent
{\it (a)} The proton spectrum is: 
\begin{eqnarray}
\lefteqn{
N_p(E_p) ~=~ N_{p,0}\, E_p^{-q_p}   \hspace{0.25cm}   {\rm cm}^{-3}\, {\rm GeV}^{-1}  
}
\label{eq:CRp_sp}
\end{eqnarray}
for $m_p c^2 < E_p < E_p^{max}$ (energies in GeV). The quantities $N_{p,0}$, $q_p$ are free parameters whose values will be 
determined by fitting the model to the data.
\smallskip

\noindent
{\it (b)} Secondary electrons are produced from $\pi^\pm$ decays following p-p interactions of the protons with ambient gas: 
their spectrum has no free parameters once the proton spectrum is specified. A summary of secondary electron production is 
given in Section A.2.2. The secondary electron spectrum will be analytically approximated as 
\begin{eqnarray}
\lefteqn{
N_{se}^{\rm fit}(\gamma) ~=~ 
N_{se,0} \gamma^{-q_1} \,\left( 1+ \frac{\gamma}{\gamma_{b1}}\right)^{q_1-q_2} e^{-\left({\gamma \over \gamma_{b2}}\right)^\eta}   
\hspace{0.25cm}   {\rm cm}^{-3}\, {\rm GeV}^{-1},  
}
\label{eq:CRe_sec}
\end{eqnarray}
where $q_1$, $q_2$ and $\gamma_{b1}$, $\gamma_{b2}$ are the low- and high-energy spectral indices and breaks; and $\eta$ gauges 
the steepness of the high-end cutoff. Their numerical values will be obtained by fitting Eq.\,\ref{eq:CRe_sec} to the actual 
spectrum (Eq.\,\ref{eq:fit_ss_se_spectrum2}) which does not have any degrees of freedom. Thus, in our treatment, these are not 
free parameters.
\smallskip

\noindent
{\it (c)} Primary electrons, required to fit the synchrotron data together with the secondary electrons, are chosen such that 
they approximate the steady-state electron spectrum following injection. Electron spectra are expressed as functions of the 
Lorentz factor, $\gamma$. 

The emission spectrum from the $\pi^0$-decay has more constraining power than a generic PL given its characteristic 'shoulder'  
at $\mincir$100 MeV and its spectral cutoff corresponding to $E_p^{max}$. Thus, as in Paper I our modelling procedure begins with 
fitting a pionic emission profile to the $\gamma$-ray data with free normalisation, slope, and high-end cutoff. We then use the 
deduced (fully determined) secondary-electron spectrum together with a (modelled) primary electron spectrum to calculate the 
combined synchrotron emission using an independently estimated value of $B$. The fit to the radio data also includes, at high 
frequencies, a thermal bremsstrahlung component computed with literature-deduced values of the gas density and temperature, 
and thermal absorption at low frequencies. Finally, the full non-thermal bremsstrahlung and Comptonised-starlight $\gamma$-ray 
yields are determined using standard emissivities (see Appendix A) and added to the pionic yield to fit the $\gamma$-ray data. 
In principle, if the resulting model exceeds the LAT data, the proton spectrum is revised accordingly and the procedure is 
repeated until convergence: in practice, for M31 and M33, the leptonic $\gamma$-ray yields turn out to be quite small compared 
to the pionic yields, so no iteration is needed.

\section{M31} 

In Ackermann et al (2017), based on 7.3 yr of {\it Fermi}-LAT data, M31 was detected as extended with a significance of $4\sigma$, 
with the favored model based on a disk of $0^\circ.4$ radius. This result provides an estimate of the LAT sensitivity to source extension 
given the LAT instrumental response, the source spectrum, the diffuse backgrounds in the region, and the available statistics. 

In Xing et al. (2023), based on 14 years of LAT data, the measured emission was resolved into two regions, but now with a statistically 
significant evidence to reveal or rule out spatial extension. It is reasonable to assume that the $\gamma$-ray source associated with 
the M31 core has an extension (radius) of (at most) $0^\circ.4$ -- otherwise it would appear as extended in Xing et al. (2023), as found 
in Ackermann et al. (2007). In this subsection, we test the hypothesis of an extended $\gamma$-ray emission in the unresolved central 
region of M31, arising from truly diffuse LH emission and possibly an unresolved population of pulsars.

%
\begin{table}
 
\caption[] {M31 and M33: Interstellar medium parameters.}
\centering 
\begin{tabular}{ l  l  l  l  l  l  l  l }
\hline
\hline
\noalign{\smallskip}
Galaxy &$n_{\rm HI}$   &   $n_{\rm H_2}$  &     $n_i$     &   $Z^2$  &           EM             &       $T_e^c$      &       $T_e^w$        \\
       &{\tiny cm$^{-3}$}&{\tiny cm$^{-3}$}&{\tiny cm$^{-3}$} &     &{\tiny cm$^{-6}$ pc}       &{\tiny $\mathrm{K}$}& {\tiny $\mathrm{K}$} \\
\noalign{\smallskip}
\hline
\noalign{\smallskip}
M31     &      292     &        0.1       &      0.04     &   1.21   &           133        &           550         &     $10^4$            \\
M33     &      1.62    &       0.14       &      0.02     &   1.21   &            64        &           500         &     $10^4$            \\

\noalign{\smallskip}
\hline
\hline
\end{tabular}
\flushleft
$^\star${\small For derivations and references of these quantities, see Appendix B.}

\end{table}

\subsection{$\gamma$-ray yields}

The average ambient proton density is $n_p = n_{\rm HI} + 2\,n_{\rm H_2} + n_i \simeq 0.53$ cm$^{-3}$ (based on values listed in Table 4). 
A pionic fit, described by the parameters $q_p$, $E_p^{\rm max}$, $u_p$ (reported in Table 5) matches the log-parabola best-fit from 
Xing et al. (2023) to the LAT data. In a fully pionic model, the unusually high value {\rm of $u_p$ stems from} the low hydrogen content 
in the central region of M31 (Fig.\,12 of Nieten et al. 2006; the 'central deficiency' noted early on by, for instance, Davies \& 
Gottesman 1970 and Emerson et al. 1974): the proton energy-loss time by p-p interactions is $\propto n_p^{-1}$, so for a given proton 
injection rate a low ambient gas density implies a high $u_p$.

\begin{figure}
\vspace{14.0cm}
\includegraphics{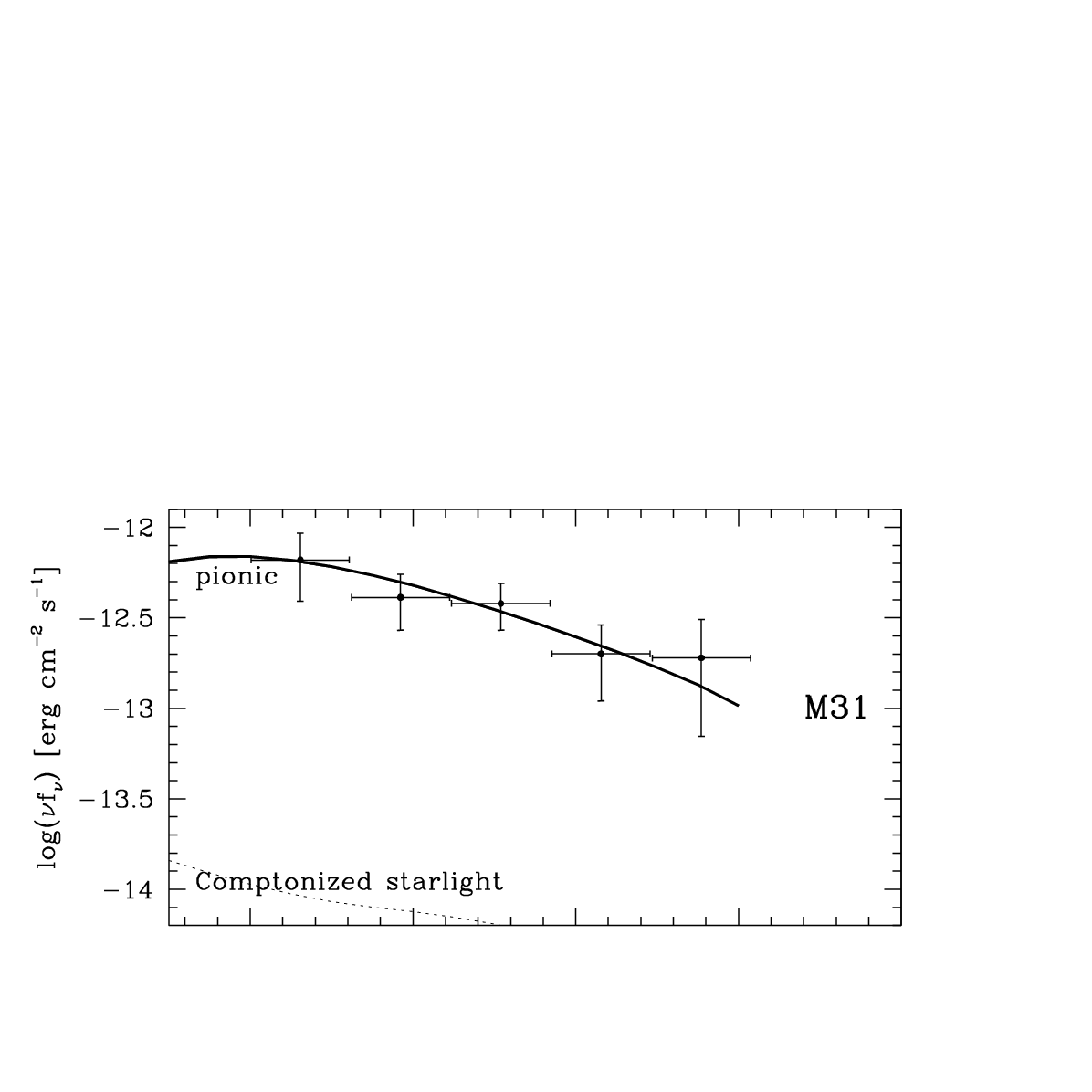}
\includegraphics{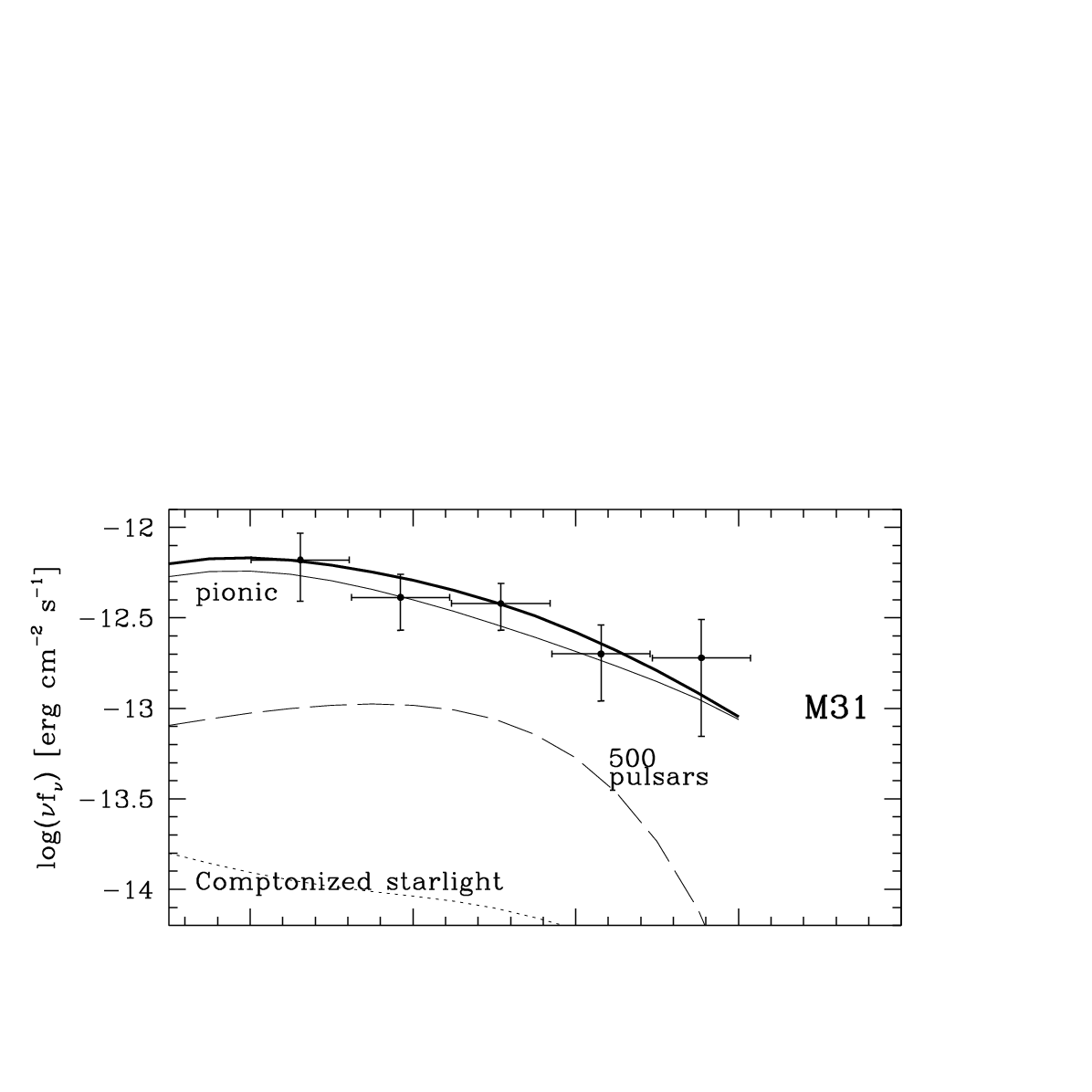}
\includegraphics{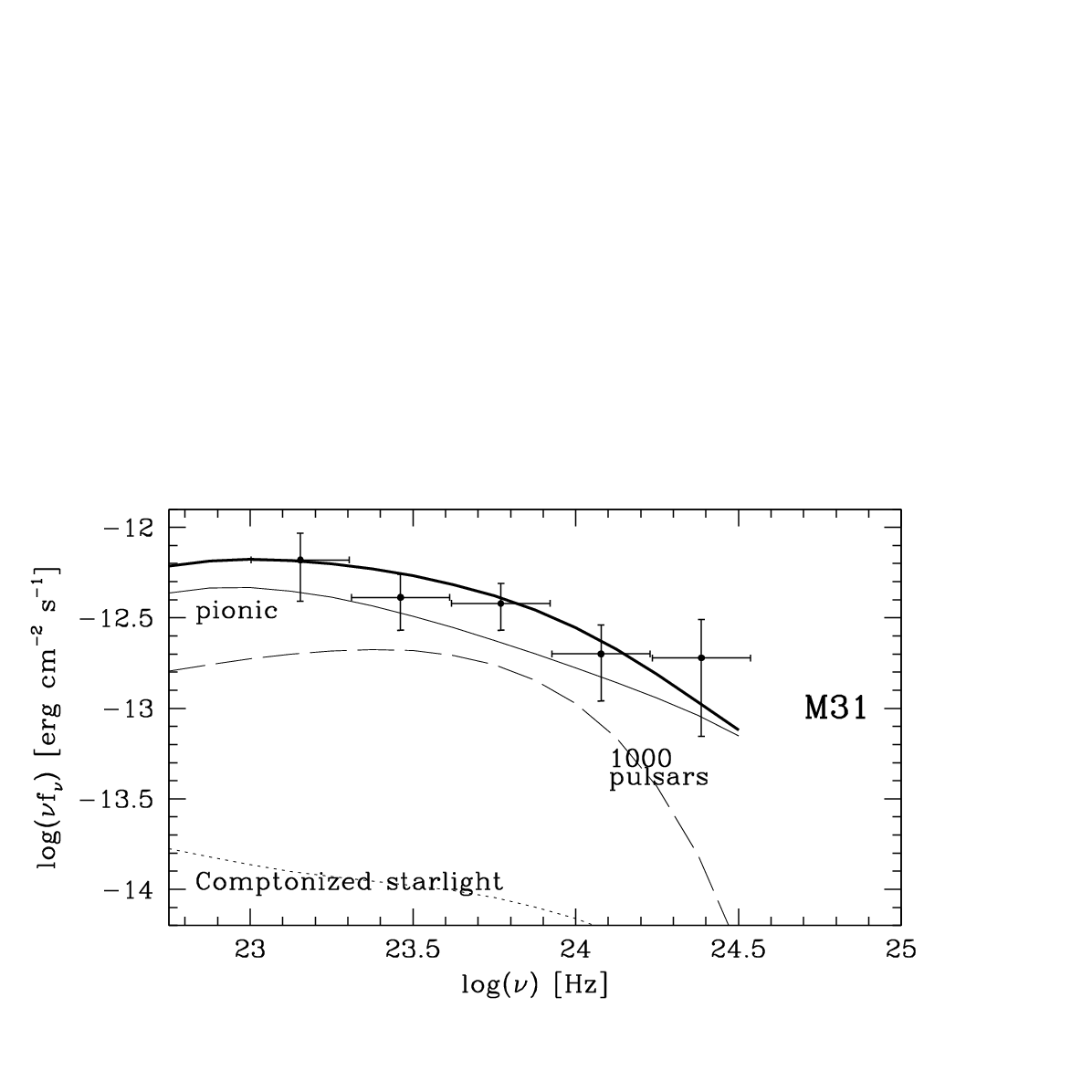}
\caption{
$\gamma$-ray spectrum of emission from the central source of M31, $R_{\star} \leq 0^{\circ}.4$. 
LAT data points (Xing et al. 2023) are shown as dots, with errorbars indicating frequency and flux 
uncertainties, and model predictions (LH: {\it top}; PSR+LH: {\it middle} and {\it bottom}) are 
shown as curves. The leptonic components used in the full calculation include primary and secondary 
electron yields. 
}
\label{fig:M31_gamma}
\end{figure}

The closely related $\pi^\pm$-decay secondary electron spectrum, $N_{se}(\gamma)$, is computed as outlined above. Once the secondary 
synchrotron radiation has been computed (assuming $B = 7.5 \pm 1.5\,\mu$G; Fletcher et al. 2004, from particles-field equipartition), 
its low-frequency residuals from the radio data are modelled as synchrotron by a PL primary electron spectrum: 
\begin{eqnarray}
\lefteqn{
N_e(\gamma) ~=~ N_{e,0}\, \gamma^{-q_e}   \hspace{0.25cm}   {\rm cm}^{-3}\, ({\rm unit ~ of}\, \gamma)^{-1}, 
}
\label{eq:CRe_PL}
.,\end{eqnarray}
in the range of $\gamma_{min} < \gamma < \gamma_{max}$ (with $\gamma_{min} >> 0$), where the quantities $N_{e,0}$, $q_e$ 
are free parameters. The ensuing primary and secondary leptonic $\gamma$-ray yields, namely, non-thermal bremsstrahlung 
and Comptonised starlight emission, are (at most) a few percent of the pionic emission. The LH $\gamma$-ray spectrum is 
plotted in Fig.\ref{fig:M31_gamma} (top).

%
%

The appropriate model for the LAT-measured $\gamma$-ray spectrum of M31 might be not just LH. The galaxy bulge and disk 
are characterised by a half-light radius of $R_e = 1$ kpc and an exponential scale length of $R_d = 5.3$ kpc (Courteau 
et al. 2011),  respectively. The upper limit to the central source radius, $R_\star = 5.45$ kpc, includes the bulge and 
$\sim$30\% of the (thin exponential) disk. So we may expect a $\gamma$-ray contribution from a fairly large unresolved 
population of $\gamma$-ray pulsars (PSR) located in the bulge and the inner disk 
\footnote{
If the LAT data need to be modelled as pure pulsar emission, a decent match is obtained for N$_{\rm psr} = 
3500$ Galactic-like pulsars emitting, through a photon spectrum $dN/dE \propto E^{-\Gamma} e^{-(E/E_{\rm 
cut})^b}$ (Abdo et al. 2013), where $\Gamma = 2$, $E_{\rm cut} = 3$ GeV, and $b = 0.6$ at a luminosity 
level $L_{>0.1\,{\rm GeV}} = 5 ~10^{34}$ erg s$^{-1}$.
}.
The $\gamma$-ray PSR in M31 are expected to be similar to those in our Galaxy. The latter average spectral parameters are 
$\Gamma = 1.6$, $E_{\rm cut} = 2.5$, and $b=1$ (Abdo et al. 2013; Lopez et al. 2018). These values define a spectral shape 
inconsistent with the M31 LAT data. We incorporate the PSR spectrum and the LH spectrum into a mixed pulsar-plus-LH (PSR+LH) 
scenario. Possible examples of PSR+LH models involve N$_{\rm psr} = 500\, (1000)$ PSR (Fig.\ref{fig:M31_gamma}, middle and 
bottom; and Table 5); since the M31 central source has $L_{>0.1\,{\rm GeV}} = 1.4 ~10^{38}$ erg s$^{-1}$ (Xing et al. 2023), 
PSR contribute 18\% (36\%) of the source luminosity in these models. Eyeball fits suggest that this family of PSR+LH models 
can nicely match the data for N$_{\rm psr} \mincir 1200$ (and $u_p \magcir 5$ eV cm$^{-3}$). If the unresolved PSR population 
is composed of msPSR, models using average Galactic msPSR $\gamma$-ray parameters, namely, $L_{>0.1\,{\rm GeV}} = 3 ~10^{33}$ 
erg s$^{-1}$, $\Gamma = 1.3$, $E_{\rm cut} = 2$, and $b=1$ (Abdo et al. 2013) can accommodate up to $\sim$$10^4$ msPSR.

\subsection{Radio yields}

The radio emission measured in Battistelli et al. (2019) was integrated over the whole galaxy. To determine the primary electron 
densities in the region appropriate for calculating the $\gamma$-ray leptonic yields within $R_{\star}$, we need to estimate 
the radio flux from this region. To do so, we should know  (in principle) the radial behavior of all the flux densities used 
in our analysis in order to build the radio SED integrated out to $R_\star$. From the literature, we took the radial profiles 
of the 408 and 842\,MHz flux densities (Gr\"ave et al. 1981, their Fig.\,9). Both profiles are point-source--subtracted and 
averaged over circular rings centred on the nucleus of M31 and have a comparable overall appearance. As the 842\,MHz profile 
appears to have a more regular, Gaussian-like shape (with a standard deviation of $\sigma \simeq 1^\circ$), we assume that it 
is an adequate representative flux density profile. Integrating over a circular $R_{\star}$ region, we estimate that $\sim$15\% 
of the total flux emanates from this region. Accordingly, our spectral fitting will be based on these reduced fluxes. Given 
the thinness of the disk, radiation produced in the outer disk is hardly captured by the $<$$R_\star$ region. 

Reducing the flux according to the 842\,MHz profile is a first-order correction to the galaxy-integrated flux. We expect 
flux-density radial profiles to steepen with increasing frequency because the corresponding higher-energy electrons diffuse 
out to shorter distances from their acceleration sites during their shorter energy loss time. 
\footnote{
The diffusion length is ${\cal L}_D = \sqrt{D \tau}$ where $D$ is the (constant) diffusion coefficient and $\tau$ is 
the particle radiative lifetime. For synchrotron losses it is $\tau = \gamma/b_{\rm syn} \propto \gamma^{-1}$ (where 
$b_{\rm syn} \propto \gamma^2$ is the synchrotron loss term, see Eq.\,\ref{eq:electr_SC}), hence, ${\cal L}_D 
\propto \gamma^{-1/2}$.
}
Moreover, the injection rate of freshly accelerated electrons is not sufficient to offset for their shorter radiative lifetimes 
because star formation rate declines fast with radius -- even faster than the gas density (based on the Schmidt-Kennicutt law 
$\Sigma \propto \sigma^N$ with 
$N > 1$).
\footnote{
$\Sigma$ and $\sigma$ are surface densities of, respectively, star formation rate and gas. 
The exact value of $N$ depends on the gas tracer (HI or H$_2$ or their sum) and the particular 
galaxy (e.g. DeGioia-Eastwood et al. 1984 for NGC\,6946, and Elia et al. 2022 for the Milky 
Way; see also reviews by Kennicutt 1998 and Kennicutt \& Evans 2012). 
}
The flux-density radial profiles at different frequencies have been measured in several nearby galaxies (e.g. M33: Tabatabaei et 
al. 2007; M51: Mulcahy et al. 2014) and a systematic (albeit usually moderate) radial steepening of the flux density profiles with 
increasing frequencies is clearly seen. { We also expect this to be the case  in M31 even though we are not aware of any available 
high-frequency data for this galaxy: if so, the emission fraction within $R_\star$ will be higher at GHz frequencies than the value 
we obtained at 842\,MHz.} So to compensate for the overcorrection inherent in the low-frequency-based first-order correction, which 
(as noted) is progressively more severe with increasing frequency, a second-order correction is needed. Lacking suitable spectral 
data, the strength of such second-order correction is uncertain: examples from M33 and M51 suggest a $\mincir$25\% increase at the 
highest frequencies. Thus, as the (unknown) second-order correction is probably small, we limited our correction to the first-order and 
calibrated the $\leq$$R_\star$ radio emission using the 842\,MHz flux density profile. The main results of this paper will not be 
appreciably affected. For example, a 25\% second-order correction can be compensated by a $\sim$40\% increase in the thermal 
free-free emission leaving the non-thermal components essentially unchanged; alternatively, a flatter ($q_e = 3$) and lower (by a 
factor of $\sim$4) electron spectrum would leave the pionic component virtually unaffected.

\begin{figure}
\vspace{9.5cm} 
\includegraphics{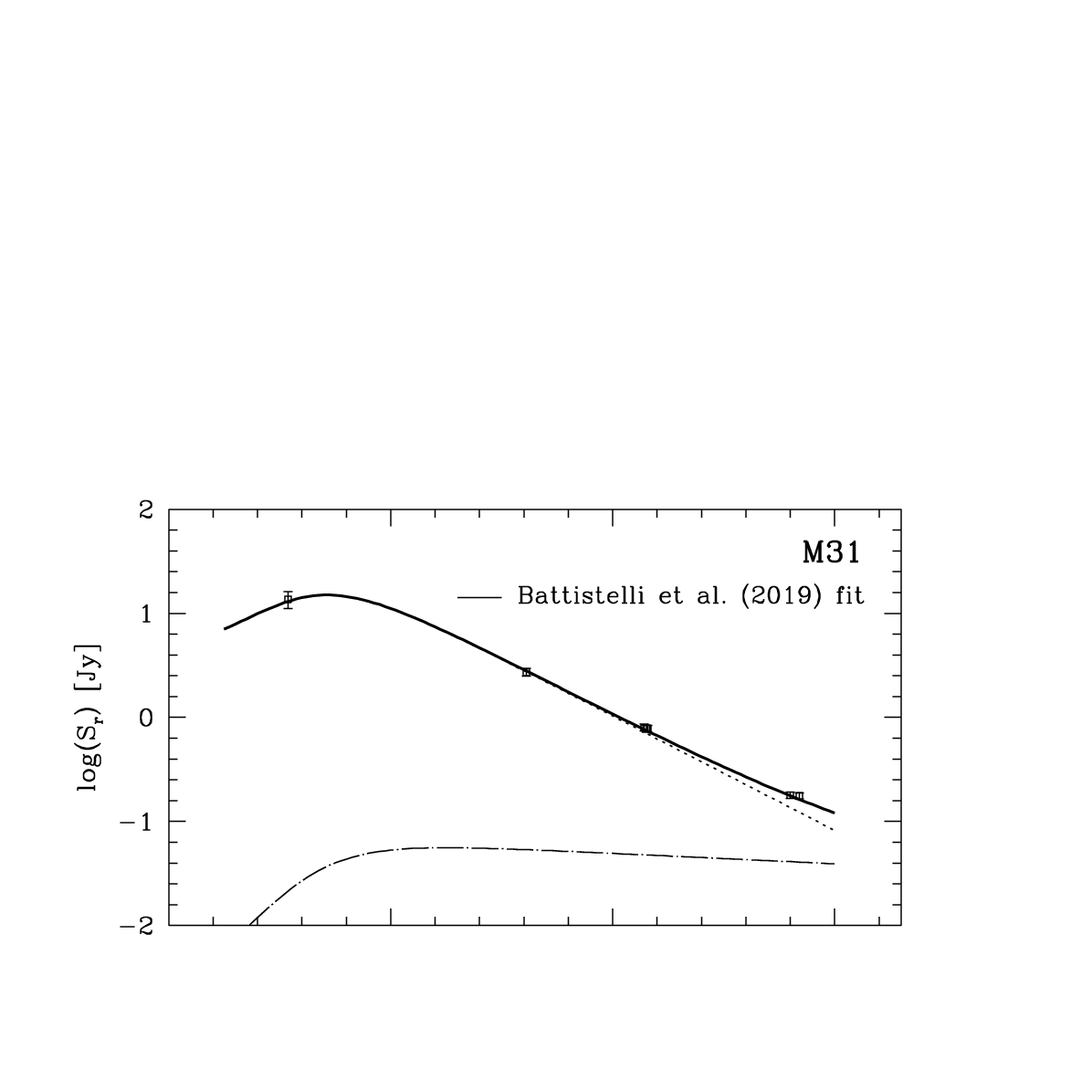}
\includegraphics{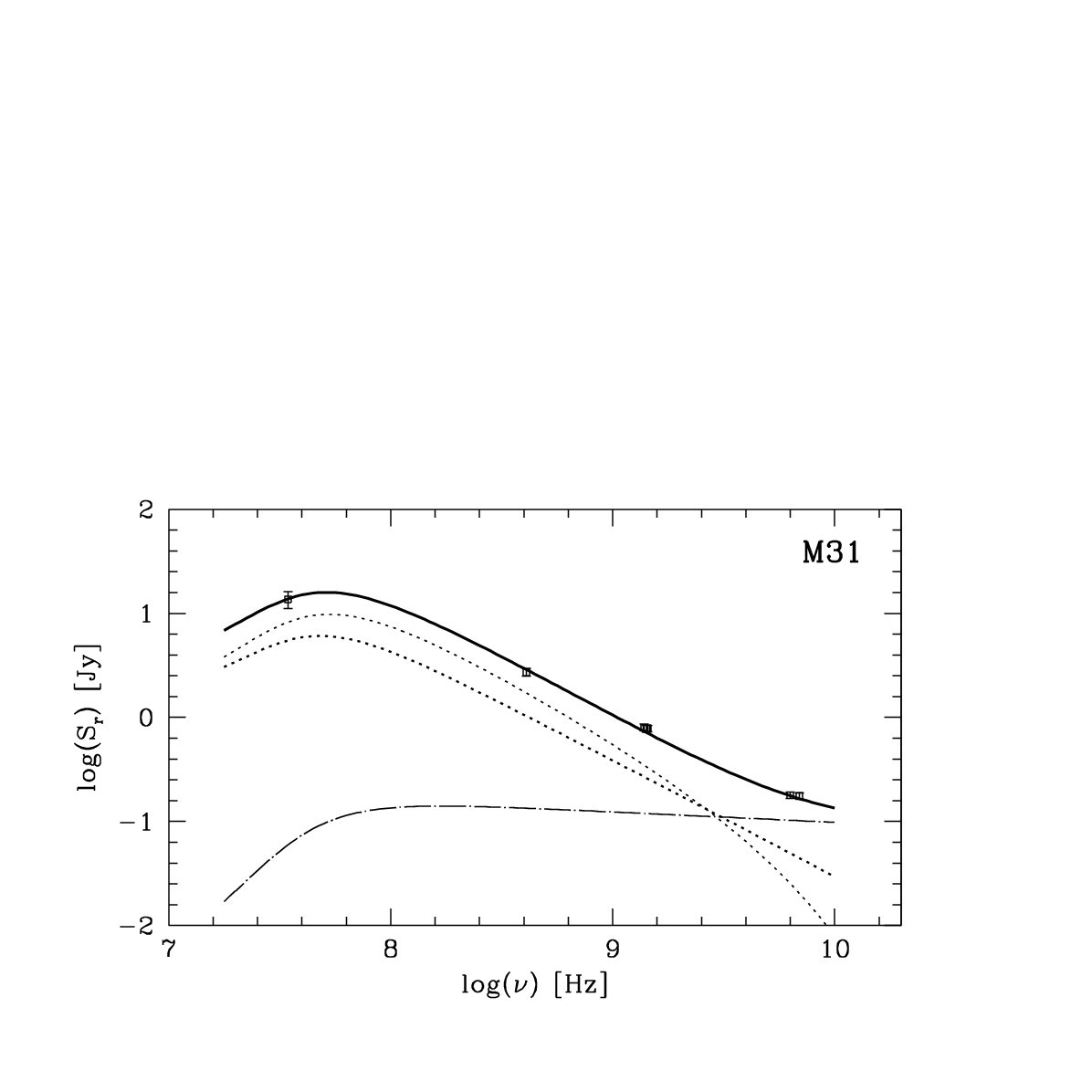}
\caption{
Radio spectrum of M31. The data points (from Battistelli et al. 2019) denote the whole-galaxy radio flux reduced to the 
estimated flux expected from within the $R_{\star}$ radial region as described in the text. Also shown are: ({\it top}) 
Battistelli et al.'s (2019) fit (single electron population synchrotron: dotted curve; thermal free-free emission: dot-dashed 
curve), and ({\it bottom}) our LH emission model. The latter includes primary and secondary synchrotron radiation (dotted 
curves of increasing flux at their peaks) and a thermal free-free component (dot-dashed curve). 
}
\label{fig:M31_radio}
\end{figure}

We wish to reproduce the M31 radio spectral model presented by Battistelli et al. (2019) self-consistently in the context of our 
approach. Synchrotron emission by secondary electrons (no degrees of freedom) dominates the radio emission up to $\sim$2 GHz. Fitting 
to the synchrotron by primary electrons results in a spectral index, $\alpha_r \simeq 1.1$, matching the total synchrotron fitting 
value of $1.1^{+0.10}_{-0.08}$ from Battistelli et al. (2019). With no evidence of a spectral cutoff in the data, we set the primary 
electron maximum energy at $\gamma_{\rm max} = 6 ~ 10^4$. The synchrotron spectrum from Battistelli et al. turns over at $\nu \sim 
48$ MHz; as we do for M33 (Section 4.2.2), we interpret this feature as absorption by a cold plasma cospatial and homogeneously mixed 
with the electrons (Spitzer 1978); this is a common feature in highly inclined disk galaxies (Israel \& Mahoney 1980). M31's 
perpendicular emission measure EM$_0 \sim 30$ cm$^{-6}$ pc (Walterbos \& Braun 1994) implies, for a galaxy inclination of 
$77^{\circ}$, a line-of-sight value of EM$ = {\rm EM}_0/{\rm cos}(77^{\circ})= 133$ cm$^{-6}$ pc. Assuming a temperature $T_e^c 
= 550\,\mathrm{K}$ for the cold plasma (similar to M33, see Section 4.2.2), from Eqs.(\ref{eq:tau1}) and (\ref{eq:tau2}), we can 
derive an optical depth, which matches the model from Battistelli at al. at $\nu < 100$ MHz. A  $\nu^{-0.1}$ component in the 
Battistelli et al. model accounts for a slight spectral concavity at $\nu > 1$ GHz. We modelled it as thermal free-free emission 
in the parametrisation reported in Eq.\,\ref{eq:TH_ff_emissivity2}, which gauges this flux with the H$\alpha$ flux assuming that 
both emissions come from the same emitting (HII) regions with the same plasma parameters (temperature, density, filling factor), 
so the measured H$\alpha$ flux may be used to predict the free-free emission. The radio spectrum is shown in Fig.\ref{fig:M31_radio}, 
overlaid with the Battistelli et al. (2019) fit ({\it top}) and our model ({\it bottom}). 

%

\subsection{Broad-band results and discussion}

\begin{figure}
\vspace{13.9cm}
\includegraphics{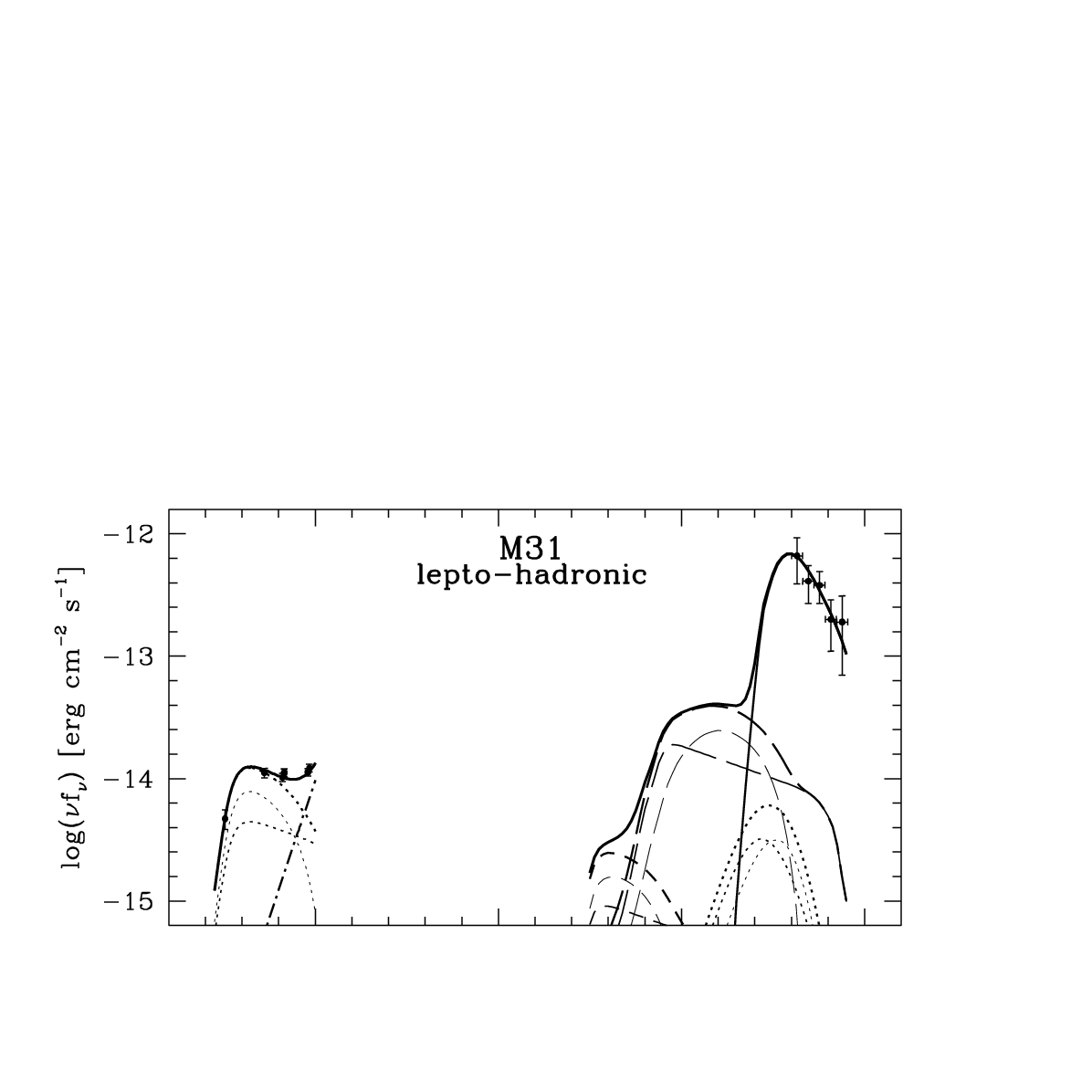}
\includegraphics{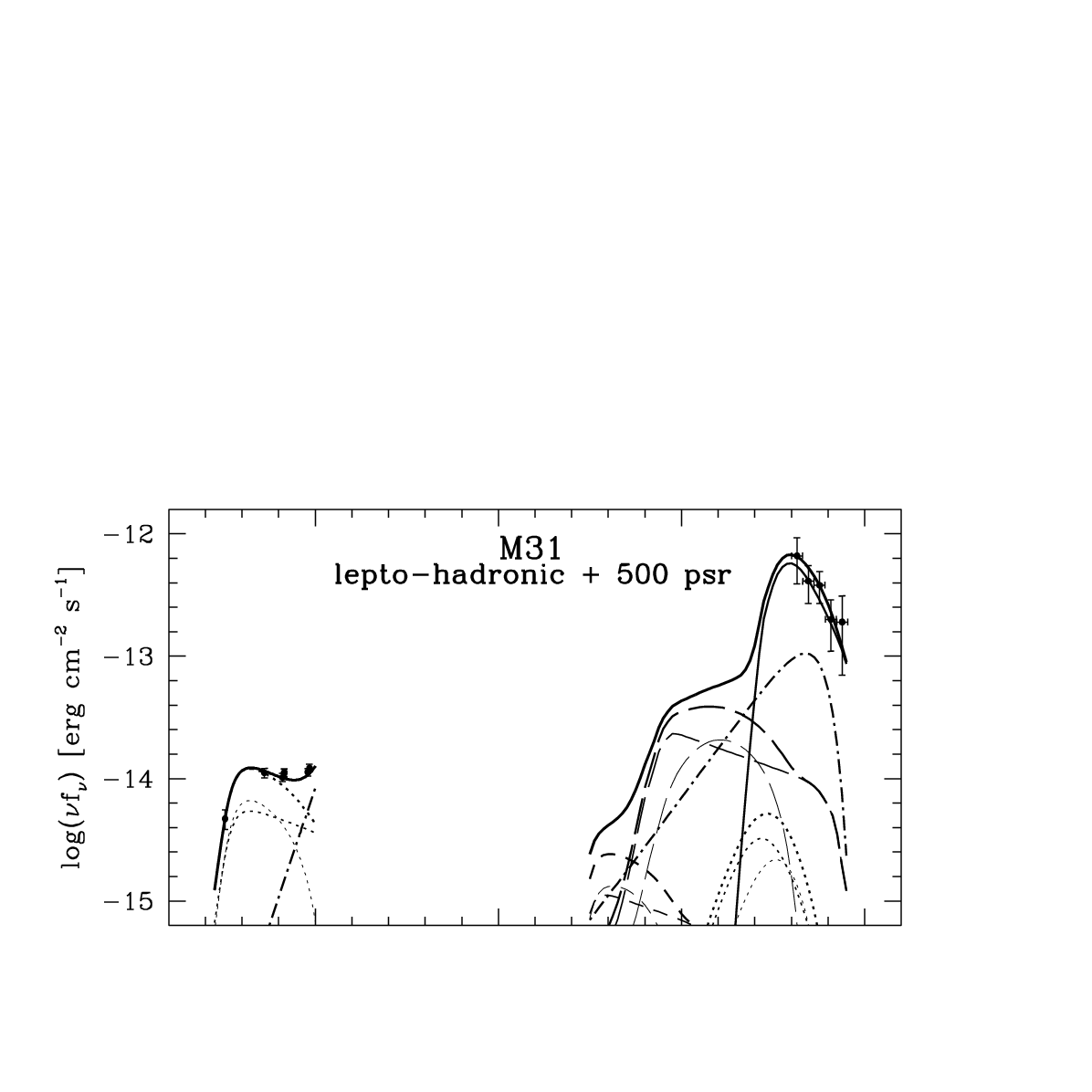}
\includegraphics{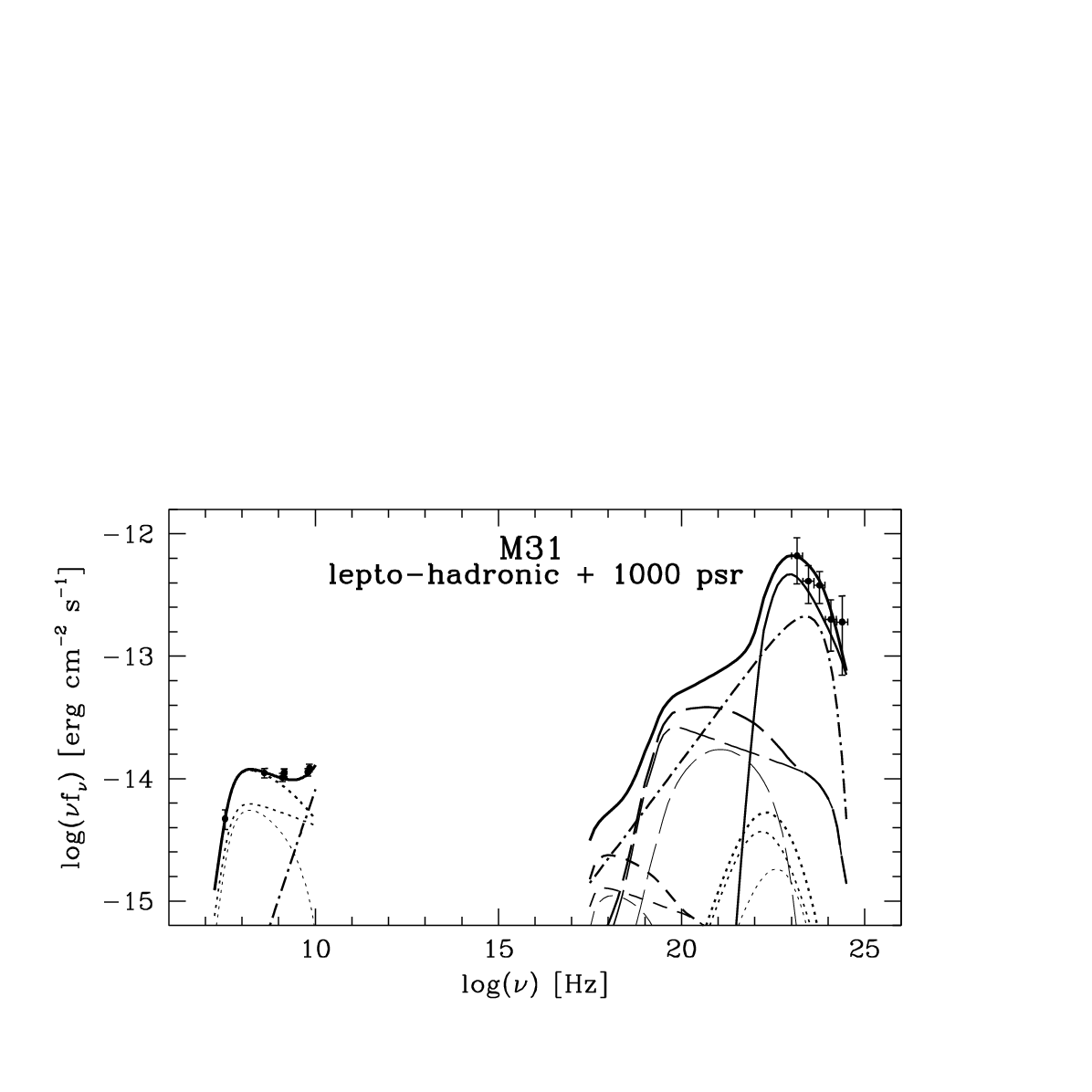}
\caption{
Broad-band lepto-hadronic SED of the M31 central source: data points are shown as dots (with errorbars indicating 
frequency and flux uncertainties), model predictions as curves ({\it top}). 
The galaxy-wide flux densities of Battistelli et al. (2019) have been renormalised to the region encompassed by 
$R_{\star}$ (i.e.$\sim$15\% of the total emission; see Section 4.1.2). Emission components are denoted by these line types: 
radio synchrotron, dotted; thermal free-free emission, dot-dashed; total radio, solid; Comptonised CMB, short-dashed; 
Comptonised starlight (EBL+FGL), long-dashed; non-thermal bremsstrahlung, dotted; pionic, solid. Leptonic emission includes 
that from primary and secondary electrons(secondary components are dominant). In the X-ray/$\gamma$-ray range, on top of the 
separate components, a thick solid line indicates the total emission. Same for a broad-band PSR+LH SED ({\it middle}). The 
500 pulsars emission component (see text) is ahown as a dot-dashed line. Same as above but for 1000 pulsars ({\it bottom}). 
Note: in this case the secondary leptonic components are subdominant.
}
\label{fig:M31_SED}
\end{figure}

The broadband SED models of the M31 central source are shown in Fig. \ref{fig:M31_SED}. The fitting parameters are given in Table 5. 
%

The small statistics of the available $\gamma$-ray data (five LAT spectral points) do not allow for a clear statement about the 
pulsar contribution, which however appears to marginally improve the fit at $\sim$2 GeV. Indeed, a consistent population of 
$\gamma$-ray pulsars is very likely to exist in M31: in the ("twin") Milky Way galaxy, about 300 pulsars have been detected by 
{\it Fermi}-LAT as of August 2023
\footnote{
https://confluence.slac.stanford.edu/display/GLAMCOG/Public+ \\
List+of+LAT-Detected+Gamma-Ray+Pulsars
} 
and thousands of unresolved msPSR have been proposed as candidates to interpret the Galactic Center "GeV excess" (Brandt \& Kocsis 
2015, Bartels et al. 2016, Arca-Sedda et al. 2018, Fragione et al 2018a,b, Miller \& Zhao 2023, Malyshev 2024; for a different 
view see Crocker et al. 2022). 

The values of $u_p$ obtained from the SED modelling are much higher than the local Galaxy value of 1 eV cm$^{-3}$ (e.g. Webber 1987). 
An independent estimate (see Section 5) suggests $u_p \sim 5$ cm$^{-3}$ for the average proton energy density within $R_{\star}$. If 
so, the PSR+LH model with 1000 PSR may be more realistic. 

We note a similarity between the values of the proton spectral index in M31, deduced from our SED modelling, and in the Milky Way, 
from the measured cosmic-ray flux at the Sun position (e.g. Gaisser 1990). The local Galactic value, $q_p = 2.7$, is understood as 
resulting from injection and diffusion, $q_p = q_{\rm inj} + \delta$ with $q_{\rm inj} \sim 2.2$ (typical of supernova remnants like 
the Crab Nebula) and $\delta \sim 0.5$ (e.g. Ptuskin 2006). 

Any current view on the nature of the central engine may change if the {\it Fermi}/LAT spatial resolution may improve, for instance, 
as a consequence of a long future operating lifetime. The central source could either be resolved into multiple sources or confirmed 
to be pointlike at the centre. As to the latter possibility, in the next paragraph we briefly discuss the possibility that the pointlike 
source at the centre is identified with M31's nuclear supermassive black hole (BH), M31$^*$.

A comparison with Sagittarius\,A$^*$ (Sgr\,A$^*$, the supermassive BH at the Galactic centre) may be useful. Cafardo \& Nemmen 
(2021) found that the centroid of the 0.1--500 GeV emission, measured by {\it Fermi}/LAT ($\sim$11 yr of data), approaches Sgr\,A$^*$ 
as the energy increases (at 10 GeV the two positions coincide within 1$\sigma$): at the Galactic Center distance, Cafardo \& 
Nemmen's (2021) point-like source has $L_{0.1-500 {\rm GeV}} = 2.6 ~10^{36}$ erg s$^{-1}$, a value consistent with the Sgr\,A$^*$ 
bolometric luminosity. The mass of M31$^*$ is $7.5~ 10^7 M_\odot$ (Ford et al. 1994), whereas that of Sgr\,A$^*$ is $4.3~ 10^6 
M_\odot$ (Ghez et al. 2008). As for the X-ray luminosities, that of M31$^*$ is $2 ~10^{36}$ erg s$^{-1}$ (Garcia et al. 2010, Li 
et al. 2009), whereas that of Sgr\,A$^*$ fluctuates between $\sim$10$^{33}$ (baseline) and $\sim$10$^{35}$ during the hr-timescale 
daily flares (Sabha et al. 2010). Thus, both M31\,A$^*$ and Sgr\,A$^*$ emit at level $\mincir$10$^{-10} L_{\rm Edd}$, with $L_{\rm 
Edd} = 1.26 ~10^{38} (M/M_\odot)$ erg s$^{-1}$ the Eddington luminosity. This suggests that both M31$^*$ and Sgr$^*$ are currently 
dormant supermassive BHs. Therefore, if the pointlike source identified by Cafardo \& Nemmen (2021) at the Galactic Centre is the 
$\gamma$-ray counterpart of Sgr\,A$^*$, with the $\gamma$ rays produced by relativistic particles accelerated by (or in proximity 
of) the BH (Cafardo \& Nemmen 2021; Ajello et al. 2021) and if a scaling relation of yet unspecified nature (Aharonian \& Neronov 
2005) exists between the $\gamma$-ray luminosity and the BH mass, then for the $\gamma$-ray counterpart of M31$^*$ we may expect 
$L_{0.1-500 {\rm GeV}} = 5 ~10^{37}$ erg s$^{-1}$; this value that corresponds to $\sim$1/2 of the luminosity of the central source 
identified by Xing et al.'s (2023). In this case, the pointlike central source in M31 may be related to M31$^*$. A study of the flux 
variability ($t_{\rm var} = R_s/c = 10^4 ~M_{\rm BH}/(10^9 M_\odot)$ s) may clarify the issue. Lacking a clear prediction on the 
$\gamma$-ray spectral signature from a supermassive BH, we generally expected the $\gamma$-ray emission of M31$^*$ to be partly pionic, 
from relativistic protons accelerated near the BH (perhaps by winds; see Ajello et al. 2021, Kimura et al. 2021) and interacting with 
the inflowing gas, and partly leptonic from the Comptonisation of the optical/X-ray radiation in the immediate proximity of the BH 
(coming from, e.g. the accretion disk; Dermer et al. 1992). If so, the BH $\gamma$-ray emission may be similar in spectral shape to 
the diffuse pionic emission tracing the (low-density) hydrogen distribution within $R_\star$. These two emission components would not 
be separated based on current LAT data, and the high proton (and ensuing secondary electron) density derived from the purely pionic 
fit may be an artifact due to spectral confusion.

\subsection{Summary on M31}

The broad-band radio--$\gamma$-ray non-thermal emission of M31 studied in this paper originates in the central ($R \leq R_\star 
\simeq 5.5$ kpc) region. This radial scale is dictated by the {\it Fermi}-LAT sensitivity to source extension within which unresolved 
(i.e. point-like) emission has been detected by Xing et al. (2023). The radio (synchrotron and thermal free-free) emission from this 
region has been estimated starting from the all-disk integrated emission, using the radial profile of the 842 MHz emission line. 

We suggest that the emission from M31's central region can be spectrally modelled as diffuse pionic with likely contributions from an
unresolved population of $\sim$1000 Galactic-like pulsars plus the nuclear BH (M31$^*$). In fact, the central 'hydrogen hole' may 
allow for emission from the enclosed pulsar population and M31$^*$ to pass through. While the Galactic-like--pulsar emission used in 
our model differs in spectral shape from the pionic emission, the (putative) BH emission may be quite similar to it hence the two 
would not be separated based on current data. Thus, the proton component of the non-thermal interstellar gas, derived from a purely 
diffuse pionic fit to current LAT data, would be biased towards high values due to contamination from M31$^*$ emission.

\begin{table*}
\caption[] {M31 SED model parameters (within $R_{\star}=5.45$ kpc).}
\centering 
\begin{tabular}{ l  l  l  l  l  l  l  l  l  l  l  l  l}
\hline
\hline
\noalign{\smallskip}
\noalign{\smallskip}
   $N_{e0}$       &$q_e$ & $\gamma_{max}$&  $u_p$ &$q_p$&$E_p^{max}$     &  $N_{se0}$              &$q_1$&$q_2$&$\gamma_{b1}$&$\gamma_{b2}$  &$\eta$& F(H$\alpha$) \\
{\small $10^{-4}$cm$^{-3}$} &      & {\small $10^4$}&{\small eV\,cm$^{-3}$}& &{\small GeV}&{\small $10^{-11}$cm$^{-3}$}&   &     &{\small $10^2$}&{\small $10^4$}& &{\small $10^{-10}$ erg/(cm$^2$s)}\\
\hline
\noalign{\smallskip}
\noalign{\smallskip}
\hline
\hline
\multicolumn{13}{c}{Lepto-hadronic} \\
\noalign{\smallskip}
\noalign{\smallskip}
 0.88 & 3.21 &  6   &   10.7   & 2.7 &    30     & 0.75 & 0.25 & 3.9 &     5.6    & 1.15 & 5 & 0.98 \\
\noalign{\smallskip}
\hline
\noalign{\smallskip}
\noalign{\smallskip}
\multicolumn{13}{c}{Lepto-hadronic plus 500 pulsars} \\
\noalign{\smallskip}
\noalign{\smallskip}
 1.08 & 3.21 &  6   &    9.0   & 2.7 &    30     & 0.60 & 0.25 & 3.9 &     5.6    & 1.15 & 5 & 0.85 \\
\noalign{\smallskip}
\hline
\noalign{\smallskip}
\noalign{\smallskip}
\multicolumn{13}{c}{Lepto-hadronic plus 1000 pulsars} \\
\noalign{\smallskip}
\noalign{\smallskip}
 1.23 & 3.21 &   6   &   7.3   & 2.7 &    30     & 0.50 & 0.25 & 3.9 &     5.6    & 1.15 & 5 & 0.83 \\
\noalign{\smallskip}
\hline
\hline
\end{tabular}
\end{table*}

\section{M33} 

Based on 11.4\,yr of LAT data, Xi et al. (2020) detected a statistically significant emission from M33. Although it peaks on the 
northeast region of the galaxy, where the giant HII region NGC\,604
\footnote{
NGC\,604 is a very active star-forming region that, within a cavity near the centre of the nebula, contains $>$200 massive blue 
stars -- some with masses exceeding 120$\,M_\odot$ and surface temperatures of $\sim$7$\,10^4\,\mathrm{K}$). The cavity itself is 
blown by the collective stellar winds from these stars, whose strong ultraviolet radiation makes the surrounding gas fluorescent. 
}
is located, the LAT signal appears to originate from the whole galaxy. This is suggested by the fact that spatially extended 
templates based on IR maps of M33 ({\it IRAS} at 60$\mu$m, and {\it Herschel} Photoconductor Array Camera and Spectrometer [PACS] 
at 160$\mu$m) give almost equally good fits as the point-source model centred at the (NGC\,604-related) best-fit location (Xi et 
al. 2020). The morphological correlation between IR and $\gamma$-ray emission suggests the latter to be of mainly pionic origin.

\subsection{Pionic yields}

The ambient proton density is $n_p \simeq 1.92$ cm$^{-3}$ (see Table 4 for densities in the different phases of the interstellar 
gas). The parameters describing the proton spectrum that matches the LAT data are listed in Table 6; in particular, the PL slope 
agrees with the values reported by Xi et al. (2020) for the the nominal 0.1--80 GeV LAT spectrum (Fig.\ref{fig:M33}). The respective 
relativistic proton energy density is $u_p = 0.46$ eV cm$^{-3}$. 

The closely related $\pi^\pm$-decay secondary electron spectrum, $N_{se}(\gamma)$, is computed as described in Section 4. The 
parameters of its analytical fitting function, $N_{se}(\gamma)$, are given in Table 6. The corresponding synchrotron emissivity 
is specified in Eq.\,\ref{eq:synchro_emissivity_exp2PL}.

\subsection{Leptonic yields}

To compute the electrons synchrotron emission we assume the relevant average total magnetic field in which primary and secondary 
electrons are immersed to be $B = 6.5\, \mu$G (Tabatabaei et al. 2008). Once the secondary electron synchrotron yield has been 
computed, its low-frequency residuals from the data are modelled by attributing them to primary electrons. 

To find the spectrum of the primary electrons, we adopt a phenomenological approach. We first try a PL spectrum but its resulting 
synchrotron emission (see Eq.\,\ref{eq:synchro_emissivity_PL2}) grossly misses the data. We then use a smooth double-PL (2PL) spectrum: 
\begin{eqnarray}
\lefteqn{
N_e(\gamma) ~=~ N_{e0} \, \gamma^{-\alpha} \,(1+\gamma/\gamma_b)^{\alpha-\beta}   \hspace{0.25cm}   {\rm cm}^{-3}\, ({\rm unit ~ of}\, \gamma)^{-1},  
}
\label{eq:CRe_2PL_generic}
\end{eqnarray}
where $N_{e0}$ is the normalisation, $\alpha$ and $\beta$ are the low/high-energy slopes, $\gamma_b$ is the break energy, and 
$\gamma_{max}$ is the cutoff energy. Comparison with the steady-state electron spectrum in the presence of radiative losses (see 
Section 4) suggests that $\alpha = q_{\rm inj} - 1$ and $\beta = q_{\rm inj} + 1$, where $q_{\rm inj}$ is the injection spectral 
index. So the adopted 2PL spectrum is finally written as 
\begin{eqnarray}
\lefteqn{
N_e(\gamma) ~=~ N_{e0} \, \gamma^{-(q_{\rm inj} - 1)} \, (1+\gamma/\gamma_b)^{-2}
   \hspace{0.25cm}   {\rm cm}^{-3}\, ({\rm unit ~ of}\, \gamma)^{-1},  \,
}
\label{eq:CRe_2PL}
\end{eqnarray}
{ where $N_{e0}$, $q_{\rm inj}$, $\gamma_b$ are free parameters.}
The corresponding synchrotron emissivity is specified in Eq.\,\ref{eq:synchro_emissivity_2PL}.

The high measured value of the H$\alpha$ emission measure (Deharveng \& Pellet 1970) and the sub-GHz turnover of the radio 
spectrum (Israel et al. 1992) suggest free-free absorption of the radio emission at low frequencies by a cool ionised plasma. 
The Galactic-absorption corrected EM in the direction to the observer (galaxy inclination is $i=57^{\circ}$) is EM$ = 64$ 
cm$^{-6}$ pc (Monnet 1971). By definition, the EM is an integral quantity, so a detailed knowledge of the diffuse 
ionised gas along the line of sight is not needed. In agreement with Israel et al. (1992), we find that, independent of the 
electron injection index, absorption of the non-thermal emission by a $T_e \sim 500\,\mathrm{K}$ plasma is required to fit 
the $\mincir 100$ MHz data. 

\begin{figure}
\vspace{10.5cm}
\includegraphics{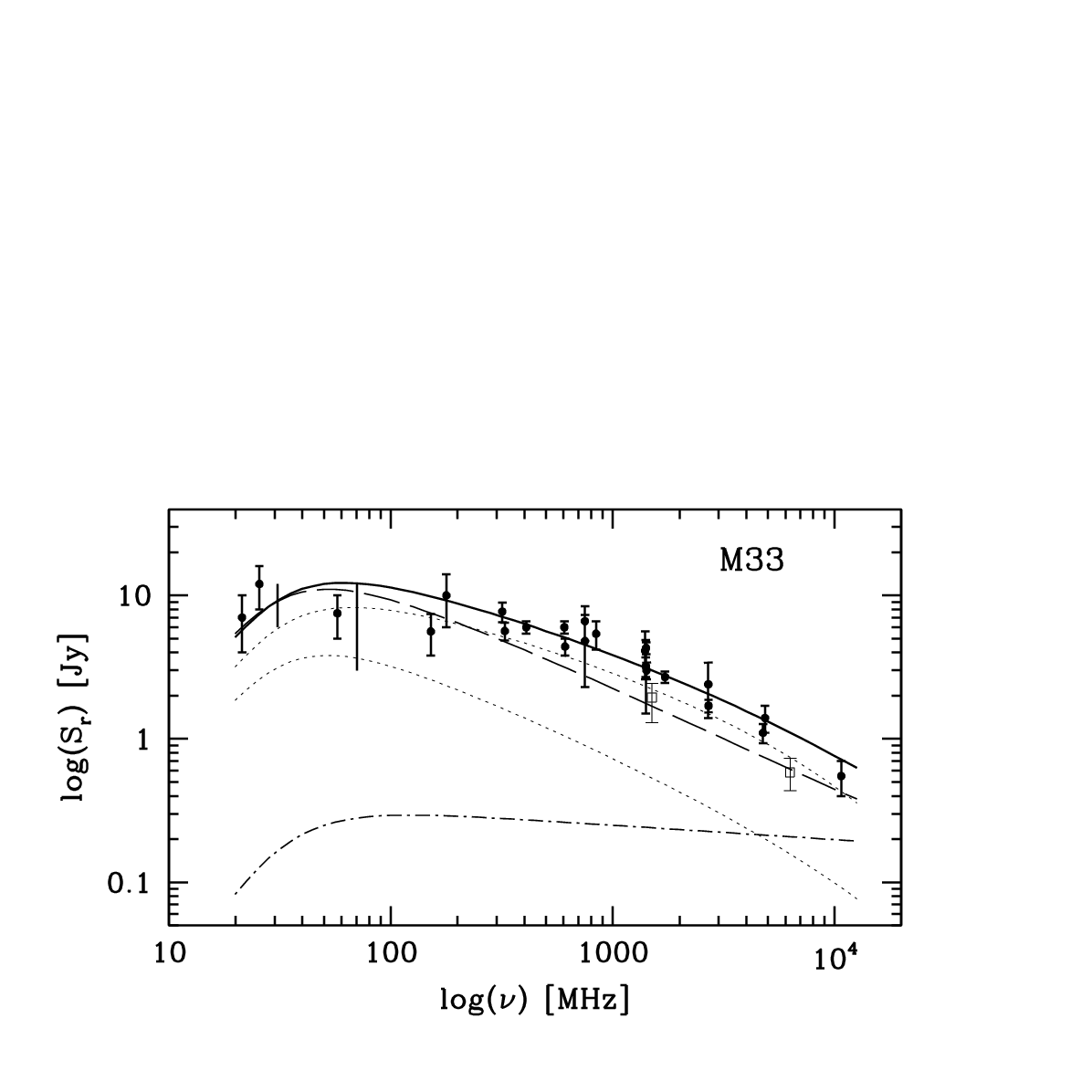}
\includegraphics{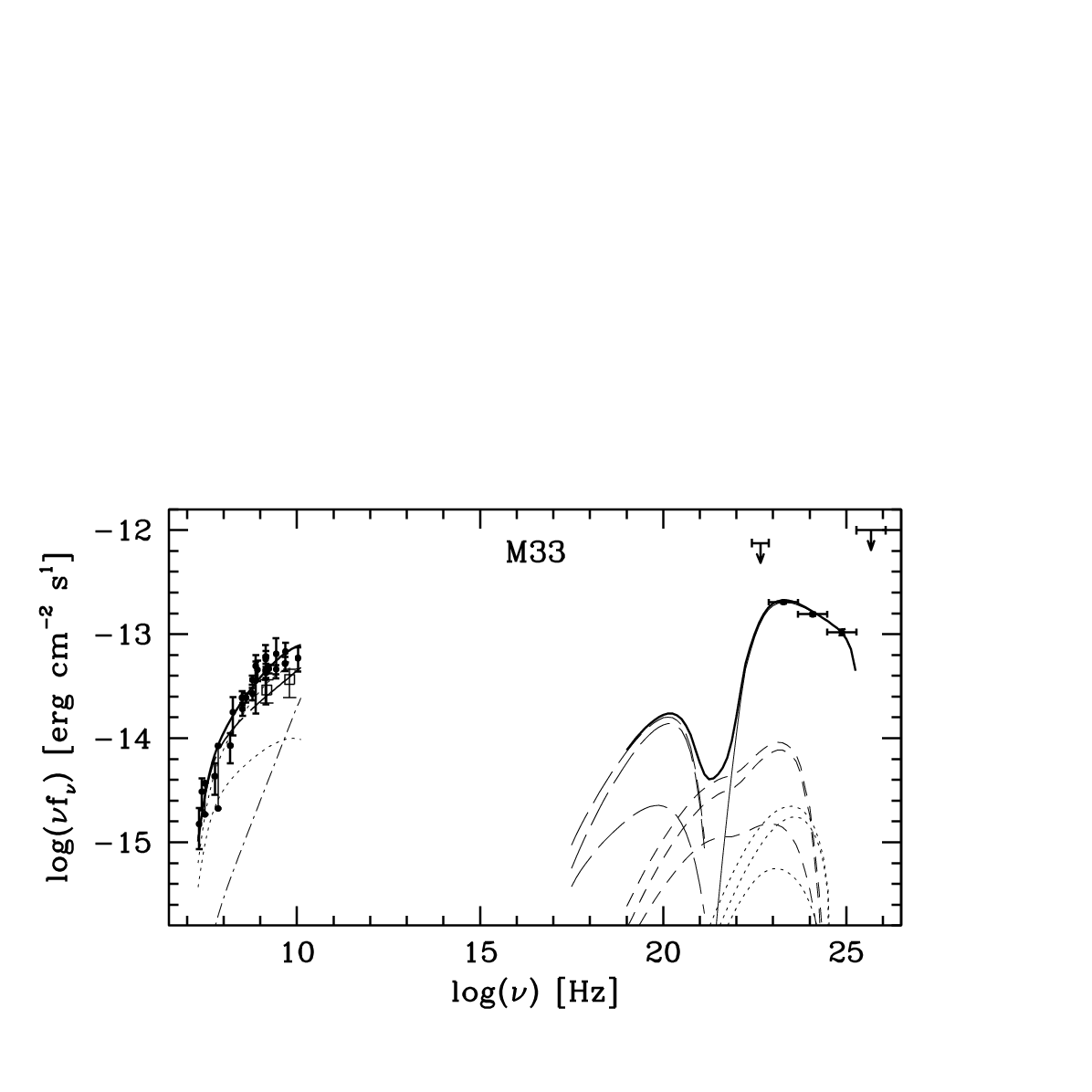}
\caption{
SED of M33. 
Radio spectrum ({\it top}): the emission model (solid curve), including primary and secondary synchrotron radiation (dotted 
curves of decreasing flux) as well as a thermal free-free component (dot-dashed curve), is overlaid with data from Table 2 (black 
dots). The empty squares represent the estimated radio flux after removing the contribution from $\leq$200 pc size sources 
(Tabatabaei et al. 2022). The total radio emission, resulting if the 'nominal' primary electron spectrum is used, is indicated 
by the dashed curve. 
Broad-band SED ({\it bottom}): data points are shown as dots, model predictions as curves. Emission components are plotted by 
the following line types: radio synchrotron, dotted; thermal free-free emission, dot-dashed; total radio, solid; Comptonised 
CMB, long-dashed; Comptonised starlight (EBL+FGL), short-dashed; non-thermal bremsstrahlung, dotted; and pionic, solid. For 
each type of non-thermal leptonic yield, the total, primary, and secondary emissions are denoted as curves of progressively 
lower flux. In the X-ray and $\gamma$-ray range, on top of the separate components, a solid line indicates the total emission. 
The empty squares and dashed curve in the radio are as described in the top panel.
}
\label{fig:M33}
\end{figure}

At high frequencies, the $\nu^{-0.1}$ component represents diffuse thermal free-free emission. This emission may be gauged to 
the H$\alpha$ flux if both come from the same HII regions since in this case the relevant warm-plasma parameters (temperature, 
density, filling factor) are the same: In this case, the measured H$\alpha$ flux may be used to predict the free-free emission. 
Following the latter approach (Klein et al. 2018), we modelled the thermal free-free emission from Eq.\,\ref{eq:TH_ff_emissivity2} 
using $T_e^w = 10^4\,\mathrm{K}$ for the {\rm warm} plasma and $F({\rm H}\alpha) = 2\, 10^{-10}$ erg cm$^{-2}$ s$^{-1}$ (only 
40\% of the integrated H$\alpha$ flux, $5\, 10^{-10}$ erg cm$^{-2}$ s$^{-1}$, is diffuse: Verley et al. 2009; also Hoopes \& 
Walterboos 2000). 
\footnote{ 
For consistency we note that computing the free-free emissivity using $Z^2n_i = n_e = 0.03$ cm$^{-3}$ (derived 
from the H$\alpha$ emission measure; Tabatabaei et al. 2008) for the warm plasma in Eq.(A9) we obtain a spectral 
flux consistent with the one derived from the integral H$\alpha$ flux in Eq.(A11). 
}
The radio measurements and model, which includes primary and secondary synchrotron and thermal emission, are shown in 
Fig.\,\ref{fig:M33} (top). 

With the electron spectra determined, we can calculate the Compton and non-thermal-bremsstrahlung yields from electrons 
scattering off, respectively, CMB-EBL-GFL photons and thermal plasma nuclei using Eqs.(A7)-(A8). Although the shape of 
the $\gamma$-ray spectrum is distinctly pionic, the Comptonised starlight peaks near the pionic peak ($\nu \sim 2\, 
10^{23}$ Hz), where it contributes 5\% of the measured flux, and the non-thermal-bremsstrahlung peaks at $\nu \sim 3\, 
10^{23}$ Hz, where it contributes 1\%. 

%
%
%

%
\begin{table*}
\caption[] {M33 SED model parameters (galaxy-wide).}
\centering 
\begin{tabular}{ l  l  l  l  l  l  l  l  l  l  l  l  l  l}
\hline
\hline
\noalign{\smallskip}
  $N_{e0}$&$q_{\rm inj}$ & $\gamma_b$& $\gamma_{max}$&  $u_p$ &$q_p$&$E_p^{max}$     &  $N_{se0}$              &$q_1$&$q_2$&$\gamma_{b1}$&$\gamma_{b2}$  &$\eta$& F(H$\alpha$) \\
  {\tiny $10^{-11}$cm$^{-3}$} &      & {\tiny $10^4$}& {\tiny $10^4$}&{\tiny eV cm$^{-3}$}& &{\tiny GeV}&{\tiny $10^{-13}$cm$^{-3}$}&   &     &{\tiny $10^2$}&{\tiny $10^4$}& &{\tiny $10^{-10}$ erg/(cm$^2$s)}\\
\noalign{\smallskip}
\hline
\noalign{\smallskip}
$3.5_{-2.0}^{+4.1}$ & $2.25 \pm 0.10$ & $1.3 \pm 0.1$  & 3.7 & 0.46 & 2.25 & 100 & 2.8 & 0.2 & 2.5 &     4    & 3 & 2.5 & 5 \\

\noalign{\smallskip}
\hline
\hline
\end{tabular}
\smallskip
\end{table*}

\subsection{Results and discussion}

While comparison of the LH SED model to the dataset is not formally a best-fit, the agreement is reasonable, given the 
independently set priors on the values of $B$ and $q_1$, and two of its features are well based. The first is the pionic 
fit to the LAT data: the $\gamma$-ray leptonic contribution is very minor, even with the appreciable level of he uncertainty 
in the electron spectral parameters (see below). The LAT data, however, show no spectral cutoff; the central value of the 
last LAT datapoint implies $E_p^{max} = 100$ GeV, and even with the uncertainty reflected in the width of the energy band, 
it is still clear that $E_p^{max} > 30$ GeV.

The second apparent feature is absorption of the synchrotron emission by warm thermal gas. It is clear that even for the 
hardest possible primary electron spectrum ($q_{\rm inj} = 2$) and in spite of the substantial scatter of data, the spectrum 
at $\leq$100 MHz can only be reproduced with absorption by a low-temperature ($500\,\mathrm{K}$) thermal gas, in accordance 
with a suggestion made by Israel et al. (1992), and similarly (for NGC\,891) by Mulcahy et al. 2018.)

The relatively flat $\gamma$-ray slope of M33 suggests that, unlike for M31, a substantial PSR contribution is unlikely. The 
PSR spectral cutoff (at several GeV) results in a spectral hump in the relevant energy range (see Fig.\,\ref{fig:M31_gamma}). 
This in contrast to the relatively flat PL-like pionic model that appears to naturally match the data. 

The uncertainty in the total (HI, H$_2$, HII) gas density affects the energy-loss coefficients used to compute the secondary 
electron spectrum (Paper I) and the ensuing normalisation and shape of the primary electron spectra and their yields. However, 
the electron density is effectively unconstrained because it depends on assuming a magnetic field value rather than using a 
value derived by modelling the non-thermal-X-ray emission as Comptonised CMB (e.g. Persic \& Rephaeli 2019a,b) or the hard 
X-ray--MeV as Comptonised infrared (IR) radiation. 

A further source of uncertainty in modelling the primary electron spectrum comes from the large scatter in the radio data. A 
strong coupling among $N_{e0},q_{\rm inj},\gamma_b$ implies a broad range of acceptable values of these parameters (Table 4) 
because a non-linear combination of $q_{\rm inj}$ and $\gamma_b$ determines the effective synchrotron slope at $\magcir$1\,GHz 
which, for an assumed $B$ (derived assuming equipartition between the energy densities of the magnetic field and relativistic 
particles; Tabatabaei et al. 2008), determines the primary electron normalisation $N_{e0}$. That said, we may surmise $q_{\rm 
inj} = 2.25$ is realistic for M33 based on the following argument. Electron indices measured in Galactic supernova remnants peak 
at $2.15 \leq q_e \leq 2.3$ (Mandelartz \& Becker Tjus 2015), and values in the same range are also found in the quietly 
star-forming Magellanic Clouds (Paper I). As for the proton indices, which closely reflect the injection index given the low 
energy losses suffered by protons via p-p interactions with the local medium, values in a similar range are found, $2.2 \leq 
q_p \leq 2.6$ (Mandelartz \& Becker Tjus 2015; Paper I). Thus, a value $q_{\rm inj} = 2.25$ is quite reasonable for both protons 
and electrons. The corresponding energy density ratio between proton and (primary) electron, $\simeq$22, is higher than its value 
for PL injection spectra, $\simeq$17 (Persic \& Rephaeli 2014), as a consequence of the higher energy losses of electrons than of 
protons. Finally, the electron cutoff energy $\gamma_{max}$ is constrained low enough as not to overproduce the flux at $\geq$2700 
GHz and high enough not to violate the highest-frequency point (interpreted as an upper limit by Israel et al. 1992). 

The data from Israel et al. (1992), while free from the major individual sources that could have been identified when the 
observations were made, are not free from the integrated source emission of fainter sources. Tabatabaei et al. (2022) pointed 
out that radio sources with sizes $\leq$200 pc constitute $\sim$36\% (46\%) of the total radio emission at 1.5 (6.3) GHz in 
the inner 4\,kpc $\times$ 4\,kpc disk of M33. Indeed, the nominal steady-state primary electron spectrum, $N_{pe}(\gamma) = 
k_{\rm inj}/(q_{\rm inj} -1)\, \gamma^{-(q_{\rm inj}-1)} / (b_0+b_1\gamma +b_2\gamma^2),$ where $\dot{N}_{\rm inj}(\gamma) = 
k_{\rm inj}\,\gamma^{-q_{inj}}$ is the electron spectral injection rate and the $b_j$ terms denote Coulomb, bremsstrahlung, 
and synchrotron and Compton energy losses, as in Eqs.\,(A23)-(A25), while neglecting diffusion and advection losses). While 
it ends up under-reproducing the Israel et al. (1992) data, it is still compatible with the suggestion from Tabatabaei et al. 
(2022). If so, the X-ray and $\gamma$-ray leptonic yields would be even lower, making the LAT-measured emission even more 
likely to be of pionic origin. 

Less noisy radio data are needed to clarify several issues. { At low frequencies more} precise $\mincir$100 MHz data would 
help to constrain the cold-plasma temperature in the range of $250\,\mathrm{K} \mincir T_e^c \mincir 2000\,\mathrm{K}$. This 
is essential because a more accurate determination of $T_e^c$ is key to our knowledge of the cold gas in galactic halos (which 
are best studied in inclined galaxies by measuring the absorption of radio emission from their disks; e.g. Dettmar 1992). For 
example, in the framework of the present analysis, the high-resolution 74 MHz VLSS 
\footnote{
VLA (Very Large Array) Low-frequency Sky Survey. 
}
data (Cohen et al. 2007) may be suitable to this aim after integrating the M33 map over the whole disk and suitably subtracting 
point sources in order to achieve consistency with the data in Israel et al. (1992). At higher frequencies, the data between 100 
MHz and 2 GHz, which is free from plasma absorption and located in the bremsstrahlung energy-loss regime
\footnote{
For the assumed $B=6.5\,\mu$G (Tabatabaei et al. 2008) the characteristic synchrotron frequency, $\nu_s = 3 \gamma^2 B$ MHz 
implies a range of electron energies $2.3\,10^3 \mincir \gamma \mincir 10^4$ emitting between 100 MHz and 1 GHz. This range 
is well inside the bremsstrahlung-loss--dominated regime, $600 \leq \gamma \leq 25000$.
},
where the electron spectrum is $N_e(\gamma) \propto \gamma^{-q_{\rm inj}}$, would constrain $q_{\rm inj}$ and thereby allow for a 
more reliable determination of the primary electron spectrum.

\section{Conclusion} 

In this paper, we present a spectral fitting of electron and proton radiative yields to radio and $\gamma$-ray measurements of 
the Local Group galaxies M31 and M33, namely: the central region (\mincir 5.5 kpc) for M31 and the whole disk for M33. For M31, 
our analysis suggests that a combination of diffuse pionic, pulsar, and nuclear-BH--related emissions could explain the LAT data. 
For M33, our analysis clearly indicates that the LAT-measured emission is mostly of pionic origin in agreement with Xi et al. 
2020). While uncertainties in the 3D galaxy structure directly affect the overall normalisations of the particles spectral 
distribution functions, the model SEDs are essentially unaffected by variations of the galaxy structure parameters.

Future measurement of diffuse non-thermal X-ray emission from the galactic disks will substantially improve the SED fits and 
result in improved determinations of the particle spectral parameters. For example, measurements of non-thermal 1\,keV flux 
that result (mostly) from Compton scattering of the electrons by the CMB fix the overall normalisation of the electron spectrum 
(Persic \& Rephaeli 2019a,b). This follows from the fact that the good pionic fit to the $\gamma$-ray emission firmly determines 
the secondary electrum spectrum, which then effectively leads to the primary electron normalisation, $N_{e0}$. Modelling the radio 
spectrum as synchrotron would then provide the (volume-averaged) magnetic field, $B$. A spectral determination of $B$ would remove 
the need to assume particle-field equipartition or to use line-averaged fields, such as those derived by Faraday rotation studies 
of extragalactic polarised sources seen through the disk. 

Given the importance of magnetic fields in modelling non-thermal SEDs, a comment on the values we have adopted in this study 
is in order. Mean values of the magnetic fields of both galaxies adopted from radio studies of M31 (Fletcher et al. 2004) and 
M33 (Tabatabaei et al. 2008) rely on the assumption of energy density equipartition between relativistic particles and magnetic 
field. In estimating relativistic electron and proton densities from radio measurements, the electron densities can be readily 
derived but assumptions must be made on the associated proton number (energy) densities that are also required to estimate 
the equipartition magnetic field, $B_{eq}$. Fletcher et al. (2004) used $u_p/u_e = 100$ in the standard equipartition formula, 
$B_{eq} \propto (1+u_p/u_e)^{2/(5+q_e)}$ (e.g. Persic \& Rephaeli 2014), whereas Tabatabaei et al. (2008) used $N_p/N_e = 100$ 
in the 'modified' equipartition formula: $B_{eq} \propto (1 + N_p/N_e)^{2/(5+q_e)}$ (Beck \& Krause 2005). Our SED model for 
M31 exhibits $u_p/u_e \sim 100$, in agreement with Fletcher et al. (2004). As for M33, the quantity $N_p/N_e$ can be estimated 
as follows. Assuming radiative losses only, particle number densities are nearly constant (when integrated over energy) over 
time, thus, the particle number ratio can be calculated at injection: for an electrically neutral non-thermal plasma: $N_p/N_e 
= (m_p/m_e)^{(q_{\rm inj}-1)/2}$ (Schlickeiser 2002). Our M33 SED model has $q_{\rm inj} = 2.25$ (Table 6), so $N_p/N_e = 110$. 
This numerical coincidences suggest that equipartition between relativistic particles and magnetic field (i.e. minimum energy 
in the non-thermal fluid; Longair 1981; Govoni \& Feretti 2004), is achieved in the central region of M31 and the disk of M33; 
these serve as a posteriori proof that our models use the $B$-values from Fletcher et al. (2004) and Tabatabaei et al. (2008) 
in a self-consistent way. Finally we note that because $B_{eq} \propto (1+\rho_p/\rho_e)^{2/(5+q_e) \simeq 0.27}$, where $\rho_j$ 
denote energy (number) density for M31 (M33), the derived equipartition fields are relatively stable against variations of these 
quantities. 

Concerning the magnetic field in M31, we should note that the approximately constant field $B \simeq 7.5\,\mu$G derived by 
Fletcher et al. (2004) spans galactocentric radii of 6\,kpc $< R <$ 14\,kpc. In the present analysis, we extrapolated that 
value inwards into the region $R < R_\star = 5.5$\,kpc relevant to our study. The justification to do so on the fact that 
the average plasma density measured out to $R \sim 10$\,kpc is $n_i \sim 0.05$ (Trudoljubov et al. 2005), based on modelling 
the interaction between supernova remnants and the diffuse hot ionised plasma with a non-equilibrium collisional ionisation 
model. This value is close to that reported in Table 4 for the central region $R \leq 5.5$\,kpc. Since $n_e = Z^2 n_i$, the 
scaling relation $B \propto n_e^{2/3}$ (Rephaeli 1988) substantiates our decision to extrapolate the Fletcher et al. (2004) 
$B$-value into the central region of M31. 

An independent estimate of $u_p$ may provide a consistency check of the preferred (mostly) hadronic models of the $\gamma$-ray 
data of M31 and M33. Combining the SN frequency with the residency timescale, which represents the characteristic proton residence 
time in the disk, $\tau_{\rm res}$, (and assuming a nominal value of the fraction of total SN energy that is channeled to particle 
acceleration) enables an estimation for $u_{\rm p}$ (e.g. Persic \& Rephaeli 2010). Assuming no appreciable mechanical energy loss 
(e.g. in driving a galactic wind), the value of $\tau_{\rm res}$ is determined by the energy-loss timescale for p-p interactions, 
$\tau_{\rm pp} = (\sigma_{\rm pp} c n_{\rm p})^{-1}$. At the most relevant energy range ($\sigma_{\rm pp} \simeq 40$\,mb), the 
characteristic value is $\tau_{\rm pp} \sim 3 ~ 10^7 n_{\rm p}^{-1}$ yr. During $\tau_{\rm res}$, a number $\nu_{\rm SN} \tau_{\rm 
res}$ of SN explode in a region of volume, ${\rm V}$, thus depositing the kinetic energy of their ejecta, $E_{\rm ej} = 10^{51}$ 
erg per SN (Woosley \& Weaver 1995), into the interstellar medium. Arguments based on the cosmic-ray energy budget in the Galaxy 
and SN statistics indicate that a fraction $\eta \sim 0.05-0.1$ of this energy is available for accelerating particles (e.g., 
Higdon et al. 1998; Tatischeff 2008). Accordingly, we express the proton energy density as: 
\begin{eqnarray}
\lefteqn{
u_{\rm p}  \simeq  {\nu_{\rm SN} \over 0.01\,{\rm yr^{-1}}}   
{\tau_{\rm res} \over 3 \times 10^7\,{\rm yr}}   {\eta \over 0.05} {E_{\rm ej} \over 10^{51}\,{\rm erg}}  \left( {{\rm V} 
\over 200\,{\rm kpc}^3} \right)^{-1}   ~ {{\rm eV} \over {\rm cm}^3},} 
\label{eq:CRp_density}
\end{eqnarray}
where the various quantities are normalised to typical Galactic values. The results are as follows:
\smallskip

\noindent
{\it (a)} In the M31 disk region, we are considering ($R \leq R_{\star} = 5.45$ kpc), the average gas density is $n_{\rm g} 
\simeq 0.5$ cm$^{-3}$ hence $\tau_{\rm res} \sim \tau_{\rm pp} = 6 ~10^7$yr. The galaxy-wide SN rate is $\nu_{\rm SN} \simeq 
0.01$ yr$^{-1}$ (type Ia: $0.006 \pm 0.004$, van den Bergh 1988; type II: 0.004, from Eq.14 of Persic \& Rephaeli 2010 using 
$L_{\rm IR}$). Assuming the local SN rate to scale as the stellar surface brightness, the SN rate integrated within $R_{\star}$ 
(corresponding to $1.09\, R_d$) is $0.3\,\nu_{\rm SN}$. Then, Eq.(\ref{eq:CRp_density}) yields $u_{\rm p} \simeq 7$ eV cm$^{-3
}$, which is compatible with the value suggested by our PSR+LH model with $10^3$ PSR. We note that a similar value, $\sim$6 eV 
cm$^{-3}$, was deduced for the inner Galactic plane from $\gamma$-ray observations (Aharonian et al. 2006).
\smallskip

\noindent
{\it (b)} In M33, the galaxy-wide average gas density is $n_{\rm p} \sim 2$ cm$^{-3}$; hence, $\tau_{\rm res} \sim \tau_{\rm 
pp} \sim 10^7$yr. The total (mostly core-collapse, Type\,II) SN rate is $\nu_{\rm SN} \simeq 0.003$ yr$^{-1}$ (van den Bergh 
\& Tammann 1991; Pavlidou \& Fields 2001). Then, Eq.(\ref{eq:CRp_density}) yields $u_{\rm p} \simeq 0.6$ eV cm$^{-3}$ (galaxy 
average), in fair agreement with the SED modelling result. 
\footnote{ 
The agreement between the values of $u_p$ given by Eq.\ref{eq:CRp_density} and those returned by the SED 
modelling also holds for the Magellanic Clouds studied in Paper I: \\
{\it (a)} Large Magellanic Cloud: $r = 3.5$ kpc, $h = 0.4$ kpc, $n_p = 0.6$ cm$^{-3}$, $\nu_{\rm SN} = 
10^{-3}$ yr$^{-1}$ (Chu \& Kennicutt 1988) $\rightarrow$ $u_p = 1.3$ eV cm$^{-3}$, versus $u_p^{\rm SED} 
= (1-1.5)$ eV cm$^{-3}$; and \\
{\it (b)} Small Magellanic Cloud: $r = 1.6$ kpc, $h = 4$ kpc, $n_p = 0.6$ cm$^{-3}$, $\nu_{\rm SN} = 6.5 
~ 10^{-4}$ yr$^{-1}$ (Kennicutt \& Hodge 1986) $\rightarrow$ $u_p = 0.6$ eV cm$^{-3}$, versus $u_p^{\rm 
SED} = 0.45$ eV cm$^{-3}$. 
}

We conclude with a remark on the detectability of the p-p related neutrino emission via $\pi^\pm$ decay from M31 and M33. 
With the proton spectrum fully determined by our fit to the measured $\gamma$-ray emission, we estimated (using Kelner 
et al.'s 2006 formalism, reported in Section A.2.3) that the broadly peaked ($\sim$0.1--10 GeV) $\pi^\pm$-decay neutrino 
fluxes from the two galaxies are $\sim$60 (M31) and $\sim$100 (M33) times lower than the flux from the Large Magellanic 
Cloud reported in Paper I. These fluxes are too low to make detections using any of the current and upcoming neutrino 
projects. 
\medskip

\noindent
{\it Acknowledgement.} 
We acknowledge constructive comments by an anonymous referee. 
MP thanks the Physics and Astronomy Department of Padova University for hospitality.


\newpage
\onecolumn

\appendix

\section{Relevant emission processes}

For completeness and convenience to the interested readers, we briefly summarise the main features of relevant radiative processes 
involving electrons and protons.

\subsection{Radiative yields of electrons and positrons}

Electrons and positrons interact with magnetic and radiation fields, emitting in a wide spectral range from radio to $\gamma$-rays. 
Relevant emission mechanisms follow below.

\subsubsection{Synchrotron radiation}

$\bullet$ Synchrotron emissivity (erg cm$^{-3}$ s$^{-1}$ Hz$^{-1}$) by the isotropically distributed PL electrons spectrum in 
Eq.\,\ref{eq:CRe_PL} is (Blumenthal \& Gould 1970, Eq.\,4.57):
\begin{eqnarray}
\lefteqn{
j_s(\nu) ~=~ { \sqrt{3} e^3 B N_{e,0} \over 4 \pi\,m_e c^2} \, \int_\Omega \sin \theta \, d\Omega_\theta \, 
\int_{\gamma_{min}}^{\gamma_{max}} \gamma^{-q_e} d\gamma \, {\nu \over \nu_c} 
\int_{\nu/\nu_c}^\infty K_{5/3}(\xi) \, d\xi      \hspace{0.5cm},       \, }
\label{eq:synchro_emissivity1}
\end{eqnarray}
where $\nu_c=\nu_0 \gamma^2 \sin \theta$ ($\nu_0 = {3eB \over 4 \pi m_ec}$ is the cyclotron frequency. The modified Bessel function 
of the second kind, $K_{\zeta}(\xi)$, has the integral representation (Watson 1922):
\begin{eqnarray}
\lefteqn{
K_{\zeta}(\xi) ~=~ \int_0^\infty e^{-\xi\, \cosh(t)} \cosh(\zeta t)\, dt \,.}
\label{eq:K53}
\end{eqnarray}
Using Eq.\,\ref{eq:K53} and setting $x = \nu/\nu_c = \nu/(\nu_0 \gamma^2 \sin \theta)$, Eq.\,(\ref{eq:synchro_emissivity1}) transforms 
into: 
\begin{eqnarray}
\lefteqn{
j_s(\nu) = N_{e,0} {\sqrt{3} \over 4} {e^3 B \over m_e c^2} \left({\nu \over \nu_0}\right)^{-{q_e-1 \over 2}} 
\int_0^\pi \sin \theta ^{q_e+3 \over 2} 
\int_{\nu \over \nu_0 \gamma_{max}^2 \sin \theta }
^{\nu \over \nu_0 \gamma_{min}^2 \sin \theta }
x^{q_e-1 \over 2}
\int_x^\infty \int_0^\infty e^{-\xi \, \cosh(t)} \cosh\left({5 \over 3} t \right)\, dt \,d\xi\, dx \, d\theta 
\,, }
\label{eq:synchro_emissivity_PL1}
\end{eqnarray}
which is finally reduced to:  
\begin{eqnarray}
\lefteqn{
j_s(\nu) ~=~ N_{e,0} {\sqrt{3} \over 4} {e^3 B \over m_e c^2} \left({\nu \over \nu_0}\right)^
{-{q_e-1 \over 2}}\int_0^\pi \sin \theta ^{q_e+3 \over 2}
\int_{\nu \over \nu_0 \gamma_{max}^2 \sin \theta }
^{\nu \over \nu_0 \gamma_{min}^2 \sin \theta } x^{q_e-1 \over 2}
\int_0^\infty e^{-x \, \cosh(t)} {\cosh\left({5 \over 3} t \right) \over \cosh\left(t \right)\,} ~
dt \,dx \, d\theta \,. }
\label{eq:synchro_emissivity_PL2}
\end{eqnarray}
\smallskip

\noindent
$\bullet$ The synchrotron emissivity by the 2PL electron spectrum in Eq.\,\ref{eq:CRe_2PL_generic} is: 
\begin{eqnarray}
\lefteqn{
j_s(\nu) ~=~ N_{se,0} \,{\sqrt{3} \over 4} \, {e^3 B \over m_e c^2}\, 
\left({\nu \over \nu_0}\right)^{-{q_1-1 \over 2}}
\int_0^\pi \sin \theta ^{q_1+3 \over 2} \,
\int_{\nu \over \nu_0 \gamma_{max}^2 \sin \theta}^{\nu \over \nu_0 \gamma_{min}^2 \sin \theta} x^{q_1-1 \over 2} \,
\left( 1 + {1\over \gamma_b} \sqrt{\nu \over \nu_0 x \sin \theta} \right)^{q_1-q_2} 
\int_0^\infty e^{-x \, \cosh(t)} {\cosh\left({5 \over 3} t \right) \over \cosh\left(t \right)\,} ~
dt \,dx \, d\theta   \,; } 
\label{eq:synchro_emissivity_2PL}
\end{eqnarray}
setting $q_1 = q_{\rm inj}-1$ and $q_2 = q_{\rm inj}+1$ yields the emissivity corresponding to the electron spectrum in Eq.\,\ref{eq:CRe_2PL}.
\smallskip

\noindent
$\bullet$ The synchrotron emissivity by the electron spectrum in Eq.\,\ref{eq:fit_ss_se_spectrum2} is: 
\begin{eqnarray}
\lefteqn{
j_s(\nu) = N_{se,0} \,{\sqrt{3} \over 4}  {e^3 B \over m_e c^2}\, \left({\nu \over \nu_0}\right)^{-{q_1-1 \over 2}}
\int_0^\pi \sin \theta ^{q_1+3 \over 2} 
\int_{\nu \over \nu_0 (\gamma_{se}^{max})^2 \sin \theta}^{\nu \over \nu_0 (\gamma_{se}^{min})^2 \sin \theta} x^{q_1-1 \over 2} 
\left( 1 + \sqrt{ \nu \over \nu_0 x \sin \theta} \, 1\over \gamma_{b1} \right)^{q_1-q_2} 
e^{ -\left({\sqrt{\nu \over \nu_0 x \sin \theta} \over \gamma_{b2}}\right)^{-\eta} }  
\int_0^\infty e^{-x \, \cosh(t)} {\cosh\left({5 \over 3} t \right) \over \cosh\left(t \right)} 
dt \,dx \, d\theta  } 
\nonumber\\
& & {} {} 
\label{eq:synchro_emissivity_exp2PL}
\end{eqnarray}
which in the present analysis describes secondary-electron synchrotron.

\subsubsection{Compton scattering}

The differential number of scattered photons in Compton scattering of electrons from the above distribution by a diluted (dilution 
factor $C_{\rm dil}$) Planckian (temperature $T$) radiation field: 
\begin{eqnarray}
\lefteqn{
n(\epsilon) ~=~ C_{\rm dil}~ {8 \pi \over h^3 c^3}~ {\epsilon^2 \over e^{\epsilon /  k_BT} -1}
\hspace{0.5cm}  {\rm cm^{-3} s^{-1} erg^{-1}} \,,}
\label{eq:phot_numb_dens1}
\end{eqnarray}
is (Blumenthal \& Gould 1970, Eq.\,2.42): 
\begin{eqnarray}
\lefteqn{
{ d^2 N_{\gamma, \epsilon} \over dt\, d\epsilon_1 } ~=~
\int_{\epsilon_{min}}^{\epsilon_{max}} 
\int_{ {\rm max} \{ {1 \over 2} \sqrt{\epsilon_1 \over \epsilon},\, \gamma_{min} \} }^
{\gamma_{max}} N_e(\gamma)\, {\pi r_0^2 c \over 2 \gamma^4} \,  {n(\epsilon) \over \epsilon^2} \,
\left( 2\epsilon_1 \, {\rm ln} {\epsilon_1 \over 4\gamma^2 \epsilon} + \epsilon_1 + 4\gamma^2 
\epsilon - {\epsilon_1^2 \over 2\gamma^2 \epsilon} \right)
~ d\gamma \, d\epsilon
\hspace{0.5cm}  {\rm cm^{-3} s^{-1} GeV^{-1}} \,,}
\label{eq:IC_emissivity2}
\end{eqnarray}
where $N_e(\gamma)$ is a generic electron spectral density (in our case, see Eqs.\,\ref{eq:CRe_PL}, \ref{eq:CRe_2PL_generic}, 
\ref{eq:fit_ss_se_spectrum2}), $\epsilon$ and $\epsilon_1$ are the incident and scattered photon energies, and $r_0 = (e^2/
m_ec^2)^2$ is the electron classical radius. In Eq.\,(\ref{eq:IC_emissivity2}), we set $\epsilon_{min}=0$ and $\epsilon_{max}/h 
= 10^{12},~ 10^{13}$, and $10^{15}$\,Hz for the CMB, IR, and optical components, respectively.

\subsubsection{Non-thermal bremsstrahlung}

The differential cross-section for emitting a photon of energy, $h \nu$, in the scattering of an electron of initial energy, $E,$ 
and final energy, $E$$-$$h\nu,$ off an unshielded static charge $Ze$ is (Blumenthal \& Gould 1970, Eq.\,3.1). The corresponding 
emissivity (in cm$^{-3}$ s$^{-1}$ erg$^{-1}$) is: 
\begin{eqnarray}
\lefteqn{
j_{\rm NT\,bremss}(\nu) ~=~ \frac{ n_i \,c\, 4 Z^2 \,\alpha\, r_0}{h\nu}  
\int_{\gamma_{min} + \frac{h\nu}{m_ec^2}}^{\gamma_{max}} 
\left[ \frac{4}{3} \left(1-\frac{h\nu}{\gamma m c^2}\right) + \left(\frac{h\nu}{\gamma mc^2}\right)^2 \right]
\left[ {\rm ln} \left( 2 \gamma \frac{\gamma mc^2-h\nu}{h\nu}\right) - \frac{1}{2} \right] \,N_{e}(\gamma) \,d\gamma 
\hspace{0.5cm} {\rm cm^{-3} s^{-1} erg^{-1}}   }
\label{eq:NT_bremss_emissivity}
\end{eqnarray}
where $n_i$ is the thermal-plasma proton density, and $\alpha=1/137$ is the Sommerfeld fine-structure constant.

\subsubsection{Thermal bremsstrahlung}

For an isotropic thermal distribution of electrons, in the optically thin case the bremsstrahlung emissivity is (Spitzer 1978, 
Eq.\,3.54): 
\begin{eqnarray}
\lefteqn{
\epsilon_{ff}(\nu) ~=~ 
\frac{32 \pi}{3} \, \left(\frac{2\pi}{3}\right)^{1/2} \, \frac{Z^2 e^6}{m^{3/2} c^3 (kT)^{1/2}} \, N_{e0} n_i \, g_{ff}(\nu,T) \, 
e^{-h\nu/kT} \hspace{0.3cm}  {\rm erg\, cm^{-3}\, s^{-1}\, Hz^{-1}}, } 
\label{eq:TH_ff_emissivity1}
\end{eqnarray}
where $g_{ff}$, which varies slowly with $\nu$ and $T$, is the Gaunt factor for free-free transitions. At radio frequencies 
($\nu << kT/h$), it is $e^{-h\nu/kT} \simeq 1$ and 
\begin{eqnarray}
\lefteqn{
g_{ff}(\nu) ~=~ 
\frac{3^{1/2}}{\pi} \, \left\{ {\rm ln} \left[ \frac{(2kT)^{3/2}}{\pi e^2 m^{1/2}} \frac{1}{\nu} \right] 
\,-\, \frac{5}{2}\, \gamma_{\rm Eul} \right\}, }
\label{eq:gaunt}
\end{eqnarray}
where $\gamma_{\rm Eul}=0.577$ is Euler's constant (Spitzer 1978, Eq.\,3.55): for radio frequencies of practical interest, $g_{ff} 
\propto T^{0.15} \nu^{-0.1}$ (Mezger \& Henderson 1967). The thermal--free-free flux may be gauged to the H$\alpha$ flux if both 
emissions come from the same emitting volume (i.e. HII regions). This is because, in this case, the relevant warm-plasma parameters 
(temperature, density, filling factor) are the same. So the measured (optical) H$\alpha$ flux may be used to predict the (radio) 
free-free emission. From Klein et al. (2018, Eq.\,17): 
\begin{eqnarray}
\lefteqn{
S_{ff}(\nu) ~=~ 
1.14 \times 10^{12} \left(\frac{T}{10^4K}\right)^{0.34} \left(\frac{\nu}{\rm GHz}\right)^{-0.1}  
\frac{F_{\rm H\alpha}}{\rm erg \,cm^{-3}\, s^{-1}}    \hspace{0.5cm}  {\rm mJy}\,, }
\label{eq:TH_ff_emissivity2}
\end{eqnarray}
which is the expression we use in our work.

\subsection{Radiative and particle yields of decaying pions}

Collisions of relativistic and ambient protons may lead, with similar probabilities, to the production of $\pi^0$ (mass $m_{\pi^0} 
= 0.135$\,GeV) and $\pi^\pm$ (mass $m_{\pi^0}=0.140$\,GeV); then, $\pi^0$ decays yield $\gamma$-rays, $\pi^0 \rightarrow 2\,\gamma$; 
secondary $e^\pm$ originate from $\pi^\pm$ decays according to the scheme $\pi^\pm \rightarrow \mu^\pm + \nu$ and $\mu^\pm 
\rightarrow e^\pm + 2\nu$ (e.g. Mannheim \& Schlickeiser 1994).

\subsubsection{Photons from $\pi^0$ decay}

The $\pi^0$-decay $\gamma$-ray emissivity is (e.g. Stecker 1971): 
\begin{eqnarray}
\lefteqn{
Q_\gamma(E_\gamma)|_{\pi^0}  ~=~ 
2 \, \int_{E^{min}_{\pi^0}}^{E^{max}_{\pi^0}} \, 
{ Q_{\pi^0}(E_{\pi^0}) \over \sqrt{ E_{\pi^0}^2 - m_{\pi^0}^2 }} \,  dE_{\pi^0} \hspace{0.5cm} {\rm cm^{-3}\, s^{-1}\, GeV^{-1}}, }
\label{eq:hadr_emissivity1}
\end{eqnarray}
where the factor of 2 accounts for the two photons emitted by the $\pi^0$ decay, 
\begin{eqnarray}
\lefteqn{
E^{min}_{\pi^0} = E_{\gamma} + \frac{m_{\pi^0}^2}{4 E_\gamma} \hspace{0.5cm} {\rm  GeV} }
\label{eq:min_pi0_energy}
\end{eqnarray}
is the minimum $\pi^0$ energy required to produce a photon of energy $E_\gamma$, $E^{max}_{\pi^0} \simeq E^{max}_p- \frac{3}{2} m_p$ 
is the maximum $\pi^0$ energy. Then, 
\begin{eqnarray}
\lefteqn{
Q_{\pi^0}(E_{\pi^0})
 ~=~ 
{4 \over 3} \pi n_p \,\int_{E^{thr}_p(E_{\pi^0})}^{E^{max}_p} \, J_p(E_p) \,
{ d\sigma (E_{\pi^0}, E_p) \over dE_{\pi^0} } \, dE_p 
\hspace{0.5cm}  {\rm cm^{-3}\,s^{-1}\,GeV^{-1}} }
\label{eq:pion_emissivity1}
\end{eqnarray}
is the spectral distribution of the neutral pions. Here, $n_p = n_i + n_{\rm HI} + 2n_{\rm H_2}$ is the ambient proton density, $J_p
(E_p) = {c \over 4 \pi} \, N_p(E_p)$ is the (isotropic) proton flux (with $N_p(E_p)$ the proton spectrum given by Eq.\,\ref{eq:CRp_sp}), 
$d\sigma (E_{\pi^0}, E_p) / dE_{\pi^0}$ is the differential cross-section for production of neutral pions with energy $E_{\pi^0}$ from 
a collision of a proton with energy $E_p$ with an ambient proton (essentially at rest), and $E^{thr}_p = m_p\,[1+({m_\pi^0}^2+4 m_p 
m_\pi^0)/(2m_p^2)] = 1.22$ GeV is the threshold energy for $m_\pi^0$ production. We applied a $\delta$-function approximation (Aharonian 
\& Atoyan 2000) to the differential cross-section by assuming that a fixed average fraction, $\kappa_{\pi^0}$ ($\simeq 0.17$ in the 
GeV-TeV region), of the proton kinetic energy, $E_p^{kin} = E_p - m_p$, is transferred to the neutral pion, namely, $d\sigma (E_{\pi^0}, 
E_p) / dE_{\pi^0} \sim \delta(E_{\pi^0} - \kappa_{\pi^0} E_p^{kin}) \, \sigma_{pp}(E_p)$, where $\sigma_{pp}$ is the inclusive total 
cross-section for inelastic p-p collisions (i.e. it contains the multiplicity of the pions produced in each hadronic interaction; 
e.g. Dermer 1986, Achilli et al. 2011, Aaij et al. 2015). The latter cross-section can be analytically approximated as (Kelner et al. 
2006, Eq.\,79)
\begin{eqnarray}
\lefteqn{
\sigma_{pp}(E_p) ~=~ \left[ 34.3 + 1.88\,{\rm ln}\left({E_p \over {\rm TeV}}\right) + 0.25\,\left[{\rm ln}\left({E_p \over 
{\rm TeV}}\right)\right]^2 \right] ~ \left[1 - \left( {E_{\rm thr} \over E_p} \right)^4 \right]^2   \hspace{0.5cm} {\rm mbarn} \,,}
\label{eq:pp_cross_section}
\end{eqnarray}
leading to 
\begin{eqnarray}
\lefteqn{
Q_{\pi^0}(E_{\pi^0})  ~=~
{1 \over 3} \, {c \over \kappa_{\pi^0}} \, n_p ~ N_p\left(m_p + {E_{\pi^0} \over \kappa_{\pi^0}}\right) ~ 
\sigma_{pp}\left(m_p + {E_{\pi^0} \over \kappa_{\pi^0}}\right) 
\hspace{0.5cm}   {\rm cm}^{-3}\, {\rm s}^{-1}\, {\rm GeV}^{-1} \,.}
\label{eq:pion_emissivity2}
\end{eqnarray}
The resulting $\gamma$-ray photon emissivity is 
\begin{eqnarray}
\lefteqn{
Q_\gamma(E_\gamma)|_{\pi^0}  ~=~ 
{2 \over 3} \, {c \over \kappa_{\pi^0}} \, n_p \, N_{p,0} \, 
\int_{E^{min}_{\pi^0}}^{E^{max}_{\pi^0}} { (m_p+E_{\pi^0} / \kappa_{\pi^0})^{-q_p} \over \sqrt{ E_{\pi^0}^2 - m_{\pi^0}^2 }} ~ 
\sigma_{pp}\left({m_p + {E_{\pi^0} \over \kappa_{\pi^0}}}\right) ~ 
dE_{\pi^0}    \hspace{0.5cm}     {\rm cm}^{-3}\, {\rm s}^{-1}\, {\rm GeV}^{-1} \,.}
\label{eq:hadr_emissivity2}
\end{eqnarray}

\subsubsection{Secondary electrons and positrons from $\pi^\pm$ decay}

The production rate of (secondary) electrons and positrons (hereafter collectively called 'electrons') is calculated from the $\pi^\pm$ 
decay rate density (e.g. Stecker 1971):
\begin{eqnarray}
\lefteqn{
Q_{\pi^\pm}(E_{\pi^\pm}) ~=~ 
{2 \over 3} c n_p \int_{E_{\rm thr}} N_p(E_p) ~ f_{{\pi^\pm}, p}(E_p, E_{\pi^\pm}) \, dE_p 
\hspace{0.5cm}  {\rm cm}^{-3}\, {\rm s}^{-1}\, {\rm GeV}^{-1} \,,}
\label{eq:ch_pion_density1}
\end{eqnarray}
where $f_{{\pi^\pm}, p}(E_p, E_{\pi^\pm})$ is the $\pi^\pm$ energy distribution for an incident proton energy $E_p$. The $\pi^\pm$ 
energy distribution can be well approximated by assuming that a constant fraction, $k_{\pi^\pm}$, of the proton kinetic energy is 
transferred to charged pions, so that 
\begin{eqnarray}
\lefteqn{
f_{{\pi^\pm}, p}(E_p, E_{\pi^\pm}) ~=~ \sigma_{pp}(E_p) ~\delta(E_{\pi^\pm} - k_{\pi^\pm} E_p^{\rm kin}) \, . }
\label{eq:ch_pion_distrib}
\end{eqnarray}
where $\sigma_{pp}(E_p)$ is the total inclusive cross-section of inelastic p-p collisions (see above). For the range of $q_p$ values of 
interest here, $k_{\pi^\pm}= 0.25$ (Kelner et al. 2006). Substituting Eq.\,(\ref{eq:ch_pion_distrib}) in Eq.\,(\ref{eq:ch_pion_density1}) 
we obtain
\begin{eqnarray}
\lefteqn{
Q_{\pi^\pm}(E_{\pi^\pm}) ~=~ {2 \over 3} \,{c \over k_{\pi^\pm}}\, n_p \, N_{p,0} 
~ \left( m_p+{E_{\pi^\pm} \over k_{\pi^\pm}} \right)^{-q_p} ~
\sigma_{pp}\Big\{m_p+{E_{\pi^\pm} \over k_{\pi^\pm}} \Big\}     \hspace{0.5cm}  {\rm cm}^{-3} {\rm s}^{-1} {\rm GeV}^{-1} \,.}
\label{eq:ch_pion_density2}
\end{eqnarray} 

Next, we calculate the source spectrum of secondary electrons, $Q_{se}(\gamma)$ (Gould \& Burbidge 1965; Ramaty \& Lingenfelter 1966). 
In the muon ($\sim$ pion) frame, the secondary-electron distribution is: 
\begin{eqnarray}
\lefteqn{
P(\tilde{\gamma}) ~=~ {2 \tilde{\gamma}^2 \over A^3} \left(3 - {2 \tilde{\gamma} \over A}\right)\,,}
\label{eq:secondary_spectrum1}
\end{eqnarray}
where $\tilde{\gamma}$, which reaches a maximum of $A=105$ (corresponding to 52 MeV), is the (secondary) electron Lorentz factor in 
the muon frame. Then, $P(\tilde{\gamma})$ can be approximated by a delta function around $\tilde{\gamma}^\star = {7 \over 10} A \sim 
70 \approx {1 \over 4} {m_{\pi^\pm}  \over m_e}$, that is: $\delta \left(\tilde{\gamma}^\star - {m_{\pi^\pm}  \over 4 m_e} \right),$ 
where $\delta(x)$ is the Dirac $\delta$-function. Since $\gamma \simeq \tilde{\gamma}^\star \gamma_{\pi^\pm}$ (Scanlon \& Milford 
1965), at high energies $E_e \simeq m_e \gamma \sim m_e \tilde{\gamma}^\star \gamma_{\pi^\pm} = {1 \over 4}\,E_{\pi^\pm}$. The 
$e^\pm$-source spectrum is then $Q_{se}(E_e) = \int Q_{\pi^\pm}(E_{\pi^\pm})\, \delta(E_e -E_{\pi^\pm}) \,dE_{\pi^\pm} = Q_{se}(E_e) 
= 4\, Q_{\pi^\pm}(4 E_e)$; combining the latter with Eqs. (\ref{eq:ch_pion_density2}) and (\ref{eq:pp_cross_section}) and expressing 
electron energies as a function of the electron Lorentz factor $\gamma$, in the laboratory frame the secondary-electron injection 
spectrum is
\footnote{
For illustration purposes, we set $\sigma_{pp} = \sigma_{pp,0}$. (It is $\sigma_{pp,0} \simeq 30$\,mbarn for 
$5 \mincir E_p/{\rm GeV} \mincir 50$.) The corresponding injection spectrum, $Q_{se}(\gamma) =  (8/3) \,m_e \, 
(c/ k_{\pi^\pm}) \, n_p\, \sigma_{pp,0} \,N_{p0}\, \left[m_p + (4 m_e/\kappa_{\pi^{\pm}}) \, \gamma \right]^{-q_p}$ 
cm$^{-3}$ s$^{-1}$ (unit of $\gamma$)$^{-1}$ valid for $\gamma > \gamma_{thr}$, coincides with the exact solution 
except for a monotonic, rather than curved, spectral shape in the range $\gamma_{thr} < \gamma \mincir 150$. 
}
\begin{eqnarray}
\lefteqn{
Q_{se}(\gamma) ~=~  4m_e \, {2 \over 3} \, {c \over k_{\pi^\pm}} \, n_p\, N_{p0}\,  
\left(m_p + \frac{4 m_e}{\kappa_{\pi^{\pm}}} \gamma \right)^{-q_p} 
\sigma_{pp}\left(m_p + \frac{4 m_e}{\kappa_{\pi^{\pm}}} \gamma \right) \hspace{0.25cm}  ... 
\hspace{0.25cm} \gamma_{thr} < \gamma < \gamma_{se}^{max} 
\hspace{1.0cm}  {\rm cm}^{-3} {\rm s}^{-1} ({\rm unit \,of\,}\gamma)^{-1}  \,.} 
\label{eq:secondary_source}
\end{eqnarray}
where $\gamma_{se}^{thr} = \frac{1}{4} \frac{m_{\pi^\pm}}{m_e} = 68.5$ and $\gamma_{se}^{max} = \frac{1}{4} \frac{m_{\pi^\pm}^{max}}{m_e} 
\simeq \frac{1}{4\, m_e} (E_p^{max} - \frac{3}{2} m_p)$. 

Finally, the corresponding steady-state distribution, $N_{se}(\gamma)$, can be calculated from the kinetic equation 
\begin{eqnarray}
\lefteqn{
{d \left[b(\gamma) N_{se}(\gamma) \right] \over d\gamma} ~=~ Q_{se}(\gamma),\, }
\label{eq:kinetic_eq}
\end{eqnarray}
where, assuming transport losses to be negligible, $b(\gamma)$ is the radiative energy loss term. In a medium that consists of 
ionised, neutral, and molecular gas with densities $n_i$, $n_{\rm HI}$, and $n_{\rm H_2}$, respectively, the loss term is written 
as $b(\gamma) = b_0(\gamma) + b_1(\gamma) + b_2(\gamma)$, where: 
\begin{eqnarray}
\lefteqn{
- b_0(\gamma) = 1.1 \cdot 10^{-12} \, \sqrt{1-\gamma^{-2}} \,
\left[ n_i \, \left( 1 - \frac{{\rm ln} \, n_i}{74.6} \right) 
+ 0.4\,(n_H+2n_{H2}) \right] \hspace{0.5cm} {\rm s}^{-1}, \,}
\label{eq:electr_excitat_loss}
\end{eqnarray}
\begin{eqnarray}
\lefteqn{
- b_1(\gamma) = 1.8 \cdot 10^{-16} \gamma \,\left[n_i + 4.5\,(n_H+2n_{H2})\right] \hspace{0.5cm} {\rm s}^{-1}, \,}
\label{eq:electr_bremss}
\end{eqnarray}
\begin{eqnarray}
\lefteqn{
- b_2(\gamma) = 1.3 \cdot 10^{-9} \gamma^2 \, (B^2 + 8\pi \rho_r) \hspace{0.5cm} {\rm s}^{-1}. }
\label{eq:electr_SC}
\end{eqnarray}
These are the loss rates by, respectively, electronic excitations in ionised gas, electron bremsstrahlung in ionised gas (strong shield 
limit), and synchrotron-Compton radiation in the ambient magnetic and radiation field energy ($\rho_r$) densities. For a discussion 
of the $b_j(\gamma)$ terms, we refer to Rephaeli \& Persic (2015). The secondary electron spectrum is then: 
\begin{eqnarray}
\lefteqn{
N_{se}(\gamma) ~=~ \frac{\int_\gamma^{\gamma_{se}^{max}} Q_{se}(\gamma) \, d\gamma}{b(\gamma)}   
\hspace{0.25cm}  ... \hspace{0.25cm} \gamma_{thr} < \gamma < \gamma_{se}^{max}  \hspace{1.0cm}  {\rm cm}^{-3}~ ({\rm unit~ of~ \gamma})^{-1}\,.}
\label{eq:ss_se_spectrum1}
\end{eqnarray}
In the radiative-yield calculations, $N_{se}(\gamma)$ can be numerically approximated as 
\begin{eqnarray}
\lefteqn{
N_{se}^{\rm fit}(\gamma) ~=~ 
N_{se,0} \gamma^{-q_1} \left( 1+{\gamma \over \gamma_{b1}} \right)^{q_1-q_2} e^{-\left( {\gamma \over \gamma_{b2}} \right)^\eta}, \,} 
\label{eq:fit_ss_se_spectrum2}
\end{eqnarray}
where $q_1$ and $q_2$ are the low- and high-frequency slopes, $b_1$ and $b_2$ are the low- and high-frequency breaks, and $\eta$ is a 
parameter describing the sharpness of the high-energy spectral cutoff.

\subsubsection{Neutrinos from $\pi^\pm$ decay}

Here, we summarise (from Kelner et al. 2006) the spectral properties of the decay neutrinos of ultrarelativistic charged pions. We set 
$x \equiv E_\nu/E_{\pi^\pm}$, $E_{\pi^\pm}^{max} = E_p^{max} - 3/2\, m_p$, and $\Theta(x)$ the Heaviside function ($\Theta(x) = 1$ if 
$x \geq 0$, and $\Theta(x) = 0$ if $x < 0$) throughout.
\smallskip

\noindent
$\bullet ~ \pi \rightarrow \mu \nu_{\mu}$. In the laboratory frame, a pion of energy $E_{\pi^\pm} >> m_\pi$ emits a $\nu_\mu$ whose energy 
appears in the range $0 \leq E_\nu \leq \lambda E_\pi$ (where $\lambda = 1 - m_\mu^2/m_{\pi^\pm}^2 = 0.427$) with constant probability. The 
energy probability distribution for a single emitted neutrino is 
\begin{eqnarray}
\lefteqn{
f_{\nu_\mu^{(1)}}\left( {E_\nu \over E_{\pi^\pm}} \right) ~=~ \frac{1}{\lambda} \,\Theta(\lambda-x) }
\label{eq:nu_mu1_distribution}
\end{eqnarray}
and the injection spectrum of pion-decay muon neutrinos is 
\begin{eqnarray}
\lefteqn{
Q_{\nu_\mu^{(1)}}(E_{\nu}) ~=~ 
2 \, \int_{E_\nu/\lambda}^{E_{\pi^\pm}^{max}} Q_\pi(E_{\pi^\pm}) \, f_{\nu_\mu^{(1)}}\left( {E_\nu \over E_{\pi^\pm}} \right) \, dE_\pi{\pi^\pm}
~=~
2 \, \int_{E_\nu/\lambda}^{E_{\pi^\pm}^{max}} Q_{\pi^\pm}(E_{\pi^\pm}) \, 
\Theta\left(\lambda- {E_\nu \over E_{\pi^\pm}}\right) \, \frac{dE_{\pi^\pm}}{\lambda E_{\pi^\pm}}
 \hspace{0.5cm}  {\rm cm}^{-3}\, {\rm s}^{-1}\, {\rm GeV}^{-1}, \,}
\label{eq:nu_mu1_spectrum}
\end{eqnarray}
\smallskip
where the factor 2 accounts for the contributions from both $\pi^+$ and $\pi^-$ pions. 

\noindent 
$\bullet ~ \mu \rightarrow e \, \nu_e \, \nu_\mu$. In the laboratory frame, the energies of the neutrinos produced through the decay 
of secondary muons appear in the range of $0 \leq x \leq 1,$ according to probability distributions given, for $\nu_\mu$ and $\nu_e$,
respectively, (setting $r \equiv 1- \lambda = 0.573$) as follows: 
\begin{eqnarray}
\lefteqn{
f_{\nu_\mu^{(2)}}(x) = g_{\nu_\mu}(x) \Theta(x-r) \, + \, \left[ h_{\nu_\mu}^{(a)}(x) + h_{\nu_\mu}^{(b)}(x) \right] \Theta(r-x) \,,}
\label{eq:nu_mu2_distribution}
\end{eqnarray}
where 
\begin{eqnarray}
\lefteqn{
g_{\nu_\mu}(x) = \frac{3-2r}{9\,(1-r)^2} \,(9 x^2 - 6\, {\rm ln}\,x - 4 x^3 -5), }
\label{eq:g_nu_mu}
\end{eqnarray}
\begin{eqnarray}
\lefteqn{
h_{\nu_\mu}^{(a)}(x) = \frac{3-2r}{9\,(1-r)^2} \,(9 r^2 - 6\, {\rm ln}\,r - 4 r^3 - 5), }
\label{eq:h_nu_mu_a}
\end{eqnarray}
\begin{eqnarray}
\lefteqn{
h_{\nu_\mu}^{(b)}(x) = \frac{(1+2r)(r-x)}{9 r^2} \,\left[ 9\, (r+x) - 4(r^2 + rx + x^2) \right] \,;}
\label{eq:h_nu_mu_b}
\end{eqnarray}
and 
\begin{eqnarray}
\lefteqn{
f_{\nu_e}(x) = g_{\nu_e}(x) \Theta(x-r) \, + 
\, \left[ h_{\nu_e}^{(a)}(x) + h_{\nu_e}^{(b)}(x) \right] \Theta(r-x) \,,}
\label{eq:nu_e_distribution}
\end{eqnarray}
where 
\begin{eqnarray}
\lefteqn{
g_{\nu_e}(x) = \frac{2}{3\,(1-r)^2} \,\left[ (1-x) \left(6\,(1-x)^2 + r\,(5+5x-4x^2) \right) + 6r\, {\rm ln}\,x \right], }
\label{eq:g_nu_e}
\end{eqnarray}
\begin{eqnarray}
\lefteqn{
h_{\nu_e}^{(a)}(x) = \frac{2}{3\,(1-r)^2} \,\left[ (1-r)\,(6-7r+11 r^2 - 4 r^3) + 6r\, {\rm ln}\,r \right], }
\label{eq:h_nu_e1}
\end{eqnarray}
\begin{eqnarray}
\lefteqn{
h_{\nu_e}^{(b)}(x) = \frac{2(r-x)}{3 r^2} \,(7r^2 - 4r^3 + 7xr - 4xr^2 - 2x^2 - 4x^2r) \,.}
\label{eq:h_nu_e2}
\end{eqnarray}
The probability distribution functions in Eqs.\,\ref{eq:nu_mu1_distribution}, \ref{eq:nu_mu2_distribution}, and \ref{eq:nu_e_distribution} 
are normalised, as per $\int_0^1 f(x)\, dx = 1$. The corresponding injection spectra for $\mu$-decay muon and electron neutrinos are, 
respectively, 
\begin{eqnarray}
\lefteqn{
Q_{\nu_\mu^{(2)}}(E_{\nu})     ~=~     2~ \int_{E_\nu}^{E_{\pi^\pm}^{max}} 
f_{\nu_\mu^{(2)}}\left( {E_\nu \over E_{\pi^\pm}} \right) \, Q_{\pi^\pm}(E_{\pi^\pm}) \, {dE_{\pi^\pm} \over E_{\pi^\pm}} 
\hspace{0.5cm}  {\rm cm}^{-3}\, {\rm s}^{-1}\, {\rm GeV}^{-1}, }
\label{eq:nu_mu2_spectrum}
\end{eqnarray}
and 
\begin{eqnarray}
\lefteqn{
Q_{\nu_e}(E_{\nu})   ~=~   2~ \int_{E_\nu}^{E_{\pi^\pm}^{max}} f_{\nu_e}\left( {E_\nu \over E_{\pi^\pm}} \right) \, 
Q_{\pi^\pm}(E_{\pi^\pm}) \, {dE_{\pi^\pm} \over E_{\pi^\pm}}    \hspace{0.5cm}   {\rm cm}^{-3}\, {\rm s}^{-1}\, {\rm GeV}^{-1} \,.}
\label{eq:nu_e_spectrum}
\end{eqnarray}
In Eqs.\,\ref{eq:nu_mu2_spectrum} and \ref{eq:nu_e_spectrum}, the factor of 2 accounts for the contributions from both the $\pi^+$ 
and $\pi^-$ pions.

\section{Notes on Table 4}

\noindent
$\bullet$ M31. Neutral and molecular gas densities are derived from HI and H$_2$ surface densities (Fig.12 of Nieten et al. 2006) 
using a constant vertical scale height ($h_s = 0.24$ kpc) crossing the galactic midplane (Brinks \& Burton 1984). The hot ionised 
plasma density is derived from X-ray measurements: $n_i \sim f^{-1/2} 10^{-2} \simeq 0.04$ cm$^{-3}$, assuming a filling factor $f 
\sim 0.1$ (Kavanagh et al. 2020). The H$\alpha$ emission measure, EM, is deduced from Walterbos \& Braun (1994). 

\noindent
$\bullet$ M33. Neutral and molecular gas densities are inferred from the corresponding masses, $M_{\rm HI} = 1.9\, 10^9 M_\sun$ and 
$M_{\rm H_2} = 3.3\, 10^8 M_\sun$ (Abdo et al. 2010), for the assumed sizes. The ionised gas density is inferred from the diffuse 
H$\alpha$ emission, $0.03 \pm 0.03$ cm$^{-3}$ (Tabatabaei et al. 2008), thermal X-ray luminosity (Schulman \& Bregman 1995) and 
temperature (Long et al. 1996), and the estimated emission region, $\sim$0.01 cm$^{-3}$. Finally, the $Z^2$, EM is from Monnet (1971). 

The quantity $Z^2 \equiv n_e/n_i$ in the hot diffuse ionised plasma, measured by Kavanagh et al. (2020) as 1.21 in 
M31, is assumed to have the same value in M33. It should be noted that the same value is also measured in the Magellanic Clouds 
(Sasaki et al. 2002). The assumed warm-plasma temperature, $T_e^w$, is standard for H\,II regions.

\end{document}